\documentclass[aps, superscriptaddress,nofootinbib, reprint]{revtex4-2}

\usepackage{xcolor}
\usepackage[%
colorlinks=true,
linkcolor=blue,
citecolor=blue,
]{hyperref}
\usepackage{booktabs}
\usepackage{rotating}
\usepackage[marginal]{footmisc}
\usepackage{graphicx}
\usepackage{dcolumn}
\usepackage{bm}
\usepackage{ulem}

\usepackage{amsmath}
\allowdisplaybreaks[4] 

\usepackage[T1]{fontenc} 
\usepackage{tensor}
\usepackage{braket}  
\usepackage{bm}      
\usepackage{upgreek}
\usepackage{extarrows}

\newcommand\MaE{\mspace{2mu}\mathrm{e}\mspace{2mu}} 
\newcommand\MaPI{\mspace{2mu}\uppi\mspace{2mu}} 
\newcommand\MaD{\,\mathrm{d}} 
\newcommand\MaI{\mathrm{i}} 

\begin{document}
\title{Gravitational Waves from Gauge Quanta Produced during Inflation}

\author{Kai-Ge Zhang}
\email{zhangkaige21@mails.ucas.ac.cn}
\affiliation{International Centre for Theoretical Physics Asia-Pacific, University of Chinese Academy of Sciences, 100190 Beijing, China}
\affiliation{Taiji Laboratory for Gravitational Wave Universe, University of Chinese Academy of Sciences, 100049 Beijing, China}

\author{Jian-Feng He}
\email{hejianfeng@itp.ac.cn}
\affiliation{Institute of Theoretical Physics, Chinese Academy of Sciences (CAS), Beijing 100190, China}
\affiliation{School of Physical Sciences, University of Chinese Academy of Sciences, No.19A Yuquan Road, Beijing 100049, China}

\author{Chengjie Fu}
\email{fucj@ahnu.edu.cn}
\affiliation{Department of Physics, Anhui Normal University, Wuhu, Anhui 241002, China}

\author{Zong-Kuan Guo}
\email{guozk@itp.ac.cn}
\affiliation{Institute of Theoretical Physics, Chinese Academy of Sciences (CAS), Beijing 100190, China}
\affiliation{School of Physical Sciences, University of Chinese Academy of Sciences, No.19A Yuquan Road, Beijing 100049, China}
\affiliation{School of Fundamental Physics and Mathematical Sciences, Hangzhou Institute for Advanced Study, University of Chinese Academy of Sciences, Hangzhou 310024, China}


\begin{abstract}
A fast-rolling axion can transfer its kinetic energy to a gauge field via the Chern-Simons coupling, leading to copious production of gauge quanta, which can act as a source of gravitational waves (GWs) with potentially observable amplitudes.
In this work, we investigate GW production in a spectator axion model when strong backreaction is taken into account. We find that decreasing the decay constant of the axion enhances GW production. Since the initial value of the axion is larger than its quantum fluctuations, such a condition imposes a lower bound on the axion dacay constant, which sets an upper bound on the amplitude of the energy spectrum of GWs. As a result, the amplitude of the predicted GW energy spectrum is lower than $10^{-10}$ in the nHz to mHz frequency range.
\end{abstract}
\maketitle


\section{Introduction}

Inflation provides a compelling extension to the standard Big Bang cosmology, addressing key issues such as the horizon, flatness, and monopole problems \cite{Guth:1980zm, Sato:1980yn, Linde:1981mu, Albrecht:1982wi, Starobinsky:1980te}. In this framework, the early universe underwent a brief period of exponential expansion, smoothing out inhomogeneities and laying down the nearly scale-invariant spectrum of primordial perturbations observed in the cosmic microwave background (CMB). The standard single-field slow-roll inflation predicts both curvature and tensor perturbations originating from quantum vacuum fluctuations, each exhibiting a nearly scale-invariant spectrum \cite{Starobinsky:1979ty, Mukhanov:1981xt, Hawking:1982cz, Guth:1982ec, Starobinsky:1982ee, Abbott:1984fp}. The tensor perturbations are commonly referred to as primordial gravitational waves (GWs). Since the tensor spectrum depends solely on the energy scale of inflation as $H^2 / M_{\mathrm{pl}}^2$, it provides unique insights into the physics of the early Universe \cite{ Lyth:1984yz, Lyth:1996im, Boubekeur:2012xn, Baumann:2006cd}.
Current constraints from CMB polarization experiments, particularly the latest Planck \cite{Planck:2018jri, Planck:2019kim, Tristram:2020wbi} and BICEP/Keck data \cite{BICEP2:2018kqh, BICEP:2021xfz}, have set stringent upper bounds on the tensor-to-scalar ratio, $r < 0.034$ at 95\% CL \cite{Tristram:2021tvh}. However, the red-tilted spectral characteristic, in conjunction with the current observational constraints at CMB scales, places it beyond the reach of both current and future GW instruments.
Nevertheless, GWs produced through alternative mechanisms, such as those sourced by amplified perturbations \cite{Senatore:2011sp, Barnaby:2012xt, Biagetti:2013kwa, Biagetti:2014asa, Mirbabayi:2014jqa, Fujita:2014oba, Yu:2023ity, Ananda:2006af, Baumann:2007zm, Kohri:2018awv, Fu:2019vqc, Domenech:2019quo, Domenech:2020kqm, Pi:2020otn, Cai:2018dig}, remain accessible to observational probes.
These GW signals represent promising targets for current and next-generation observatories operating across a broad frequency range, from nano-Hertz (nHz) to kilo-Hertz (kHz), such as EPTA \cite{EPTA:2023fyk, EPTA:2023gyr, EPTA:2023xxk}, NANOGrav \cite{NANOGrav:2015aud, NANOGRAV:2018hou}, SKA \cite{Carilli:2004nx, Janssen:2014dka}, LISA \cite{LISA:2017pwj}, Taiji \cite{Ruan:2018tsw}, LIGO \cite{Harry:2010zz}, and Virgo \cite{VIRGO:2014yos}.

To generate detectable GWs sourced from amplified perturbations, special mechanisms, such as tachyonic instability or parametric resonance, are typically required \cite{Barnaby:2010vf, Sorbo:2011rz, Meerburg:2012id, Urban:2013spa, Barnaby:2011qe, Zhou:2020kkf, Peng:2021zon}. Among the various models proposed, Chern-Simons coupling $\chi F \tilde{F}$ during inflation has emerged as a particularly promising candidate. This mechanism exploits the coupling between a rolling axion-like field and a gauge field, where the axion-like field acts either as the inflaton or as a spectator \cite{Barnaby:2010vf, Sorbo:2011rz, Barnaby:2011vw, Barnaby:2011qe, Cook:2011hg, Dimopoulos:2012av, Linde:2012bt, Meerburg:2012id, Urban:2013spa, Mukohyama:2014gba, Namba:2015gja, Garcia-Bellido:2016dkw, Agrawal:2017awz, Ozsoy:2017blg, Agrawal:2018mrg, Papageorgiou:2019ecb, Campeti:2022acx, Ozsoy:2024apn, He:2024bno, Alaei:2025ryv}. As demonstrated in multiple studies, the axion's transient rolling motion induces tachyonic instability in one helicity mode of the gauge field, leading to exponential amplification of its fluctuations. These excited gauge fields subsequently source chiral GWs through quadratic terms in the energy-momentum tensor. Since the excited gauge quanta are massless, their quadrupole moment is not suppressed as it would be if sourced by massive particles \cite{Barnaby:2012xt}. To generate GWs at small scales without violating constraints from large-scale observations, these models typically introduce a brief fast-roll phase, usually lasting less than a few {\it{e}}-foldings, during the middle of inflation \cite{Ferreira:2014zia, Ozsoy:2017blg}. By appropriately selecting the inflationary potential, the generated GWs can reach the sensitivity of current or future detectors, while simultaneously maintaining consistency with existing observational constraints at large scales \cite{Giare:2020vhn, Campeti:2020xwn, LISACosmologyWorkingGroup:2022jok, Garcia-Bellido:2023ser,  NANOGrav:2023hvm, Figueroa:2023zhu, Unal:2023srk, Niu:2023bsr, Dimastrogiovanni:2023juq, Corba:2024tfz, Maiti:2024nhv, Alam:2024fid}.

On the other hand, the produced gauge quanta can backreact on the axion field through the Chern-Simons coupling. If the backreaction is strong, it decelerates the axion and suppresses both particle and GW production \cite{Domcke:2020zez, Peloso:2022ovc, He:2025ieo, Garcia-Bellido:2023ser}.
Typically, achieving sensitivity required by next-generation detectors necessitates the production of a substantial amount of gauge quanta, rendering backreaction effects non-negligible.
To accurately study the dynamics in the strong backreaction regime, numerical techniques are essential \cite{Cheng:2015oqa, Notari:2016npn, DallAgata:2019yrr, Gorbar:2021rlt, Durrer:2023rhc, vonEckardstein:2023gwk, Iarygina:2023mtj, Garcia-Bellido:2023ser, Caravano:2022epk, Galanti:2024jhw, Caravano:2024xsb, Figueroa:2024rkr, Sharma:2024nfu, Lizarraga:2025aiw, Bhattacharya:2025guc}.
Our previous work has shown that increasing the maximum amplitude of the GW spectrum requires enhancing the slope of the potential while keeping the coupling constant moderately small \cite{He:2025ieo}. This paper extends our previous analysis to a more realistic scenario, in which the axion is treated as a spectator field with a cosine potential. Such a potential commonly arises in string theory-inspired axion models \cite{Banks:2003sx, Arkani-Hamed:2006emk, Svrcek:2006yi, Silverstein:2008sg, McAllister:2008hb, Kobayashi:2015aaa, CaboBizet:2016uzv, DallAgata:2019yrr, Dimastrogiovanni:2023juq}.
To enhance GW production, a sharper axion potential is required. However, if the potential is too sharp, the distance between the initial value of the axion field and the peak of the potential becomes smaller than the quantum fluctuation. We find that, due to this constraint, the maximum achievable slope of the potential is limited, which in turn bounds the maximal amplitude of the generated GWs.

This paper is organized as follows.
In Section \ref{sec: dynamics}, we describe the action employed in this work, which consists of an inflaton with an E-model potential, a spectator axion-like field with a cosine potential, and a gauge field coupled to the axion field via a Chern-Simons term $\frac{\alpha}{4 f_{\mathrm{a}}} \chi \tilde{F}^{\mu\nu} F_{\mu\nu}$.
We then discuss the dynamics of this system and examine the influence of model parameters, with particular emphasis on the strong backreaction regime. In Section \ref{sec: gws}, we numerically study the maximum GW energy spectrum as a function of the parameter $f_{\mathrm{a}}$, finding that a smaller $f_{\mathrm{a}}$ can slightly enhance the spectrum.
In Section \ref{sec: ini_cond}, we analyze how the initial value of the axion field relates to the peak position of the GW spectrum. We also derive a constraint on the minimum allowable value of $f_{\mathrm{a}}$ by requiring that the distance between the initial axion value and the peak of the potential be larger than the quantum fluctuation.

Throughout this paper, we adopt natural units with $\hbar = c = 1$, and define the reduced Planck mass as $M_{\mathrm{pl}} \equiv (8\pi G)^{-1/2}$. The symbol $t$ denotes cosmic time, while $\tau$ denotes conformal time. A dot, as in $\dot{\phi} \equiv \mathrm{d} \phi / \mathrm{d} t$, indicates a derivative with respect to cosmic time, and a prime, as in $\phi' \equiv \mathrm{d} \phi / \mathrm{d} \tau$, indicates a derivative with respect to conformal time.

\section{Dynamics}
\label{sec: dynamics}

In this work, we consider a realistic model consisting of an inflaton, a spectator axion, and a gauge field coupled to the axion via a Chern-Simons term. The action is given by
\begin{eqnarray}
  S &=& \int \! \MaD^4x \, \sqrt{-g} \biggl(
    \frac{M_{\mathrm{pl}}^2}{2} R
    - \frac{1}{2} \partial^\mu \phi \partial_\mu \phi
    - V(\phi)\nonumber
    - \frac{1}{4} F^{\mu\nu} F_{\mu\nu}^{\phantom{\mu\nu}} \\
    && - \frac{1}{2} \partial^\mu \chi \partial_\mu \chi - U(\chi)
    - \frac{\alpha}{4f_{\rm a}} \chi F^{\mu\nu} \widetilde{\vphantom{\nu}F}_{\mu\nu}
  \biggr).
\label{eq: L_axion}
\end{eqnarray}
where $\phi$ is the inflaton, $\chi$ is a axion-like field, $F_{\mu\nu} \equiv \partial_{\mu} A_{\nu} - \partial_{\nu} A_{\mu}$ is the field strength of a $U(1)$ gauge field and $\tilde{F}^{\mu\nu} \equiv \frac{1}{2 \sqrt{-g}} \eta^{\mu\nu\alpha\beta} F_{\alpha\beta}$ is its dual, with the totally antisymmetric tensor $\eta^{\mu\nu\alpha\beta}$ defined such that $\eta^{0123} = 1$.
The parameter $f_{\rm a}$ is the decay constant of the axion $\chi$ and has mass dimension one, while the parameter $\alpha$ is dimensionless and controls the coupling strength.
Throughout this paper, we use the spatially flat Friedmann-Lematre-Robertson-Walker (FLRW) metric $\mathrm{d}s^{2} = -\mathrm{d}t^{2} + a^{2}(t)\mathrm{d} \bm{x}^{2} = a^{2}(\tau) (-\mathrm{d}\tau^{2} + \mathrm{d}\bm{x}^{2})$.
In this paper, we assume that the inflaton drives a phase of slow-roll inflation, with its potential chosen to be consistent with current CMB observations. A representative example is the potential from the E-model class of $\alpha$-attractor models \cite{Carrasco:2015pla, Kallosh:2022feu}, given by $V(\phi) = V_{0} \left[1 - \exp\left( - \sqrt{ \frac{2}{3 \alpha_{\mathrm{att}}}} \phi \right) \right ]^{2}$, with $V_{0} = 1.578 \times 10^{-10} \, M_{\mathrm{pl}}^4$ and $\alpha_{\mathrm{att}} = 1.8$.
The potential of the axion takes the following form,
    \begin{equation}
    \label{eq: Uchi}
    U(\chi) = \frac{\Lambda^{4}}{2}
    \left[ 1 + \cos\left( \frac{\chi}{f_{\rm a}} \right) \right],
    \end{equation}
    which commonly arises in string-inspired axion models \cite{Banks:2003sx, Arkani-Hamed:2006emk, Svrcek:2006yi, Silverstein:2008sg, McAllister:2008hb, Kobayashi:2015aaa, CaboBizet:2016uzv, DallAgata:2019yrr, Dimastrogiovanni:2023juq}. We set the parameter $\Lambda$ such that $\Lambda^{4} = 0.01 V_{0}$, thereby ensuring that the axion remains a spectator and does not affect the inflationary dynamics.

From the Lagrangian Eq. \eqref{eq: L_axion} combined with the spatially flat FLRW metric, we can obtain the background equations,
\begin{align}
  \label{eq: a_eof_1}
  & H^{2} = \frac{1}{3 M_{\mathrm{pl}}^2} \rho, \\
  \label{eq: a_eof_2}
  & \frac{\ddot{a}}{a} + \frac{1}{2} \left( \frac{\dot{a}}{a}
  \right)^{2}
  = - \frac{1}{2 M_{\mathrm{pl}}^{2}} P, \\
  & \ddot{\phi} + 3 H \dot{\phi} + V_{,\phi} = 0, \\
  \label{eq: chi_eof}
  & \ddot{\chi} + 3 H \dot{\chi} + U_{,\chi} = \frac{\alpha}{f_{\rm a}}
  \braket{\bm{E} \cdot \bm{B}}.
\end{align}
Here, $\bm{E}$ and $\bm{B}$ denote the electric and magnetic fields associated with the gauge field, defined by \cite{Barnaby:2010vf, Sorbo:2011rz, Garcia-Bellido:2023ser}
\begin{align}
  & E_{i}(t) = - \dot{A}_{i} / a, \\
  & B_{i}(t) = \epsilon_{ijk} \partial_{j} A_{k} / a^{2}.
\end{align}
The angle bracket $\langle \cdots \rangle$ in Eq. \eqref{eq: chi_eof} denotes the ensemble average of the fields, which can be computed by evaluating the vacuum average of the corresponding field operators. We will give their explicit expressions in the next section. The averaged terms encode the backreaction of the quantum field on a classic background. The total energy density $\rho$ and pressure $P$ are given by
\begin{align}
\label{eq: rho_tot}
  \rho =&
  \frac{1}{2} \dot{\phi}^{2}
  + \frac{1}{2} \dot{\chi}^{2}
  + V(\phi) + U(\chi) + \frac{1}{2}\braket{ \bm{E}^{2} + \bm{B}^{2} }
 , \\\label{eq: p_tot}
  P =& \frac{1}{2} \dot{\phi}^{2} +  \frac{1}{2} \dot{\chi}^{2}
  - V(\phi) - U(\chi) + \frac{1}{6}\braket{\bm{E}^{2} + \bm{B}^{2}}.
\end{align}

From the action \eqref{eq: L_axion}, the gauge field action in the Coulomb gauge $A_{0}(\tau, \bm{x}) = 0$ can be expressed as
\begin{align}
  S_{\text{EM}} = \frac{1}{2} \int \! \mathrm{d}\tau \, \mathrm{d}^3 \bm{x} \Biggl(
  A_{i}'A_{i}' - \partial_{j}A_{i} \partial_{j}A_{i}
  + \frac{\alpha}{f_{\rm a}} \chi' \varepsilon_{ijk}
  A_{i} \partial_{j}A_{k} \Biggr),
\label{eq: L_Aq}
\end{align}
We then decompose gauge field in the Fourier space as
\begin{align}
  & \hat{A}_{i}(t, \bm{x})
  =  \int \frac{\mathrm{d}^{3} \bm{k}}{(2 \MaPI)^{3 / 2}}
  \MaE^{ \MaI \bm{k} \cdot \bm{x} } \hat{A}_{i}(t, \bm{k}) \nonumber \\
  \label{eq: Ak_decompose}
  & =  \int \frac{\mathrm{d}^{3} \bm{k}}{(2 \MaPI)^{3 / 2}}
  \sum_{\lambda=\pm} \MaE^{ \MaI \bm{k} \cdot \bm{x} }
    \epsilon^{\lambda}_{i}(\hat{k})
    A^{\lambda}(t, \bm{k}) \hat{a}^{\lambda}(\bm{k}) + \text{h.c.},
\end{align}
where $\lambda = +, -$ denotes the polarization index, $\hat{k} \equiv \bm{k}/k$ is a unit vector with the same direction of the wave vector $\bm{k}$, and $k \equiv |\bm{k}|$ is the norm. Here, $\epsilon_{i}^{\lambda}(\bm{k})$ is the circular polarization vector basis that satisfies
\begin{align}
  & k_{i} \epsilon^{\pm}_{i}(\hat{k}) = 0,
  ~ \varepsilon_{ijk} k_{j} \epsilon_{k}^{\pm}(\hat{k})
  = \mp \MaI k \epsilon^{\pm}_{i}(\hat{k}), \\
  & \epsilon_{i}^{\pm}(\hat{k}) \epsilon_{i}^{\pm}(\hat{k}) = 0,
  ~ \epsilon_{i}^{\pm}(\hat{k}) \epsilon_{i}^{\mp}(\hat{k}) = 1.
\end{align}
$\hat{a}_{\lambda}(\vec{k})$ and $\hat{a}_{\lambda}^{+}(\vec{k})$ are annihilation and creation operators satisfying standard commutation relations,
\begin{align}
  & [ \hat{a}_{\lambda}(\bm{k}), \hat{a}_{\lambda'}^{\dagger}(\bm{k}') ]
  = \delta_{\lambda\lambda'} \delta^{(3)}\left( \bm{k} - \bm{k}' \right), \\
  & [ \hat{a}_{\lambda}^{\dagger}(\bm{k}), \hat{a}_{\lambda'}^{\dagger}(\bm{k}') ]
  = [ \hat{a}_{\lambda}(\bm{k}), \hat{a}_{\lambda'}(\bm{k}') ] = 0.
\end{align}
Varying the action Eq. \eqref{eq: L_axion} with respect to $\left( A^{\lambda}(\tau,\bm{k}) \right)^{\!*}$ yields the following equation of motion (EoM) in Fourier space \cite{Sorbo:2011rz,Garcia-Bellido:2023ser},
\begin{equation}
  \label{eq: Aq}
  A^{\pm ''}(\tau,\bm{k}) + \left( k^{2} \mp
  \frac{\alpha}{f_{\rm a}} k \chi'(\tau) \right) A^{\pm}(\tau,\bm{k}) = 0.
\end{equation}
The initial condition of the mode function $A^{\lambda}(\tau,\bm{k})$ is determined by the Bunch-Davies vacuum,
\begin{equation}
   \left. A^{\pm}(\tau, \bm{k}) \right|_{- k \tau \gg 1}
  = \frac{1}{\sqrt{2k}} \MaE^{ - \MaI k \tau }.
\end{equation}
The term $({\alpha}/{f_{\rm a}})k\chi'(\tau)$ in Eq. \eqref{eq: Aq} arises from the Chern-Simons interaction. Without loss of generality, we focus on the case $\chi' < 0$ in this work. When the term $(k^{2}+({\alpha}/{f_{\rm a}})k\chi'(\tau))$ become negative, the mode function $A^{-}(\tau,\bm{k})$ undergoes tachyonic instability and grow exponentially \cite{Barnaby:2011vw, Ozsoy:2020ccy}. Consequently, the amplified polarization component $A^{-}(\tau,\bm{k})$ serves as the main source of GWs. This parity violation is also imprinted on the sourced GWs \cite{Seto:2007tn, Gluscevic:2010vv, Smith:2016jqs, Domcke:2019zls}.  Using the definitions of $\bm{E}$ and $\bm{B}$, one can then compute the ensemble averages in Eqs. \eqref{eq: chi_eof}, \eqref{eq: rho_tot} and \eqref{eq: p_tot} as
\begin{align}
  \label{eq: ensemble_EB}
  & \braket{\bm{E} \cdot \bm{B}} =
  - \frac{1}{4 \MaPI^{2} a^{4}}
  \sum_{\lambda=\pm} \lambda \int_{0}^{\infty}
  \mathrm{d}k k^3
  \frac{\mathrm{d} }{\mathrm{d} \tau} | A^{\lambda}(\bm{k}) |^{2}, \\
  \label{eq: ensemble_EE}
  & \braket{E^{2}} = \frac{1}{2 \MaPI^{2} a^{4}}
  \sum_{\lambda=\pm} \int_{0}^{\infty}\mathrm{d} k k^{2}
  \left| \frac{\mathrm{d}A^{\lambda}(\bm{k})}{\mathrm{d} \tau} \right|^{2}, \\
  \label{eq: ensemble_BB}
  & \braket{B^{2}} = \frac{1}{2 \MaPI^{2} a^{4}}
  \sum_{\lambda=\pm} \int_{0}^{\infty} \mathrm{d}k k^{4}
  |A^{\lambda}(\bm{k})|^{2}.
\end{align}
To numerically solve the background equations \eqref{eq: a_eof_1}-\eqref{eq: chi_eof} and the gauge field equation \eqref{eq: Aq}, we implemented a C program utilizing the Runge-Kutta algorithm. In our program, we evolve the background quantities alongside multiple discrete gauge modes $A^{\lambda}(\tau,\bm{k})$ with varying momenta $k$ simultaneously. At each iteration, we first compute the integrals specified in Eqs. \eqref{eq: ensemble_EB}-\eqref{eq: ensemble_BB} with rectangular method, where only these modes which experience tachyonic instability are used. Moreover, we evolve the gauge field with the 2nd-order Runge-Kutta algorithm and the background with the 4th-order Runge-Kutta algorithm.

Generally, the evolution can be divided into three stages: (i) Initially, the axion rolls from the peak of the potential. During this stage, the dynamics are well described by the slow-roll approximation. Since the axion moves slowly, the production of gauge quanta remains weak throughout this phase. (ii) As the axion leaves the slow-roll regime, it enters a fast-roll phase. In this stage, the gauge quanta begin to grow exponentially, and the backreaction becomes increasingly significant. (iii) Once the backreaction becomes comparable in magnitude to the slope of the potential, the system enters the strong backreaction region. The backreaction decelerates the axion, resulting in suppressed particle production and a subsequent weakening of the backreaction. As the axion accelerates again, this cycle repeats, leading to an oscillatory axion velocity.

\begin{figure}[tbp]
  \includegraphics[width=.48\textwidth]{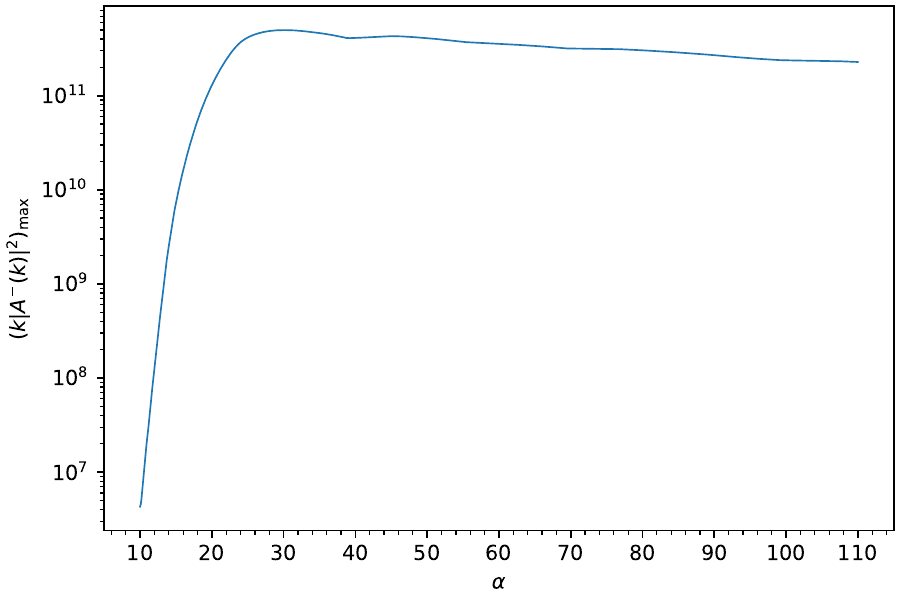}
  \caption{ Maximum value of the spectrum $k|A^{-}(k)|^{2}$ at the end of inflation as a function of $\alpha$ in the case of $f_{\rm a} = 0.08$. }
  \label{fig: Aq_peak}
\end{figure}

In our previous study \cite{He:2025ieo}, we found that for a linear potential in the strong backreaction regime, increasing the slope of the potential enhances gauge quanta production, whereas increasing the coupling constant suppresses it. This behavior arises because strong backreaction occurs when the slope term in Eq.~\eqref{eq: chi_eof} becomes comparable to the backreaction term, leading to the condition
\begin{equation}
\label{eq: strongbr_cond}
\tilde{\alpha} \braket{\bm{E} \cdot \bm{B}}
\sim U_{,\chi} = - \frac{\Lambda^{4}}{2 f_{\rm a}} \sin\left(\frac{\chi}{f_{\rm a}}\right),
  \end{equation}
  where we define $\tilde{\alpha} \equiv \alpha / f_{\rm a}$ and treat it as an independent variable with respect to $f_{\rm a}$. Since $\braket{\bm{E} \cdot \bm{B}}$ characterizes particle production, satisfying the condition \eqref{eq: strongbr_cond} implies that increasing ${\Lambda^{4}}/{(2 f_{\rm a})}$ enhances particle production, while increasing $\tilde{\alpha}$ suppresses it.
  The influence of $\alpha$ is illustrated in Fig. \ref{fig: Aq_peak}, which shows the maximum value of the spectrum $k|A^-(k)|^{2}$ at the end of inflation as a function of $\alpha$. For $\alpha \lesssim 30$, the system lies in the weak backreaction regime, and the maximum value of the spectrum $k|A^-(k)|^{2}$ increases exponentially. Beyond this threshold, the system enters the strong backreaction regime, and the maximum value becomes slightly suppressed.

From Eq. \eqref{eq: strongbr_cond}, it may appear that increasing $V_{0}$, and thus $H^{2}$, would enhance the production of gauge quanta and consequently lead to stronger GW signals. However, this is not necessarily the case. According to Eq. \eqref{eq: ensemble_EB}, the factor $k/a$ effectively reduces to $H$ near horizon crossing, which provides the dominant contribution to the integral. Therefore, increasing $V_{0}$ also increases this factor, which in turn suppresses particle production. On the other hand, the GW power spectrum in Eq. \eqref{eq: sourced_tensor_spectrum} contains a factor of $H^{2}$, so increasing $V_{0}$ directly amplifies the amplitude of the GW spectrum. Our numerical results show that these two effects approximately cancel each other, resulting in a GW spectrum that remains nearly unaffected by changes in $V_{0}$.

\section{Amplitude of the GW energy spectrum}
\label{sec: gws}

The tensor perturbation of the FLRW metric are
\begin{equation}
\label{eq: ds^2}
\mathrm{d}s^{2} = a^{2}(\tau) \left[
    -\mathrm{d}\tau^{2} + \left( \delta_{ij} + h_{ij}(\tau, \boldsymbol{x}) \right) \mathrm{d}x_{i} \mathrm{d}x_{j}
\right],
\end{equation}
where $h_{ij}$ is traceless $h_{ii} = 0$ and transverse $\partial_{j} h_{ij} = 0$.
The EoM of $h_{ij}$ obeys
\begin{equation}
\label{eq: hij_eom}
h_{ij}'' + 2 \frac{a'}{a} - \Delta h_{ij}
= \frac{2}{M_{\mathrm{pl}^{2}}}
\tensor{{\Pi}}{_{ij}^{lm}} T_{lm}^{\mathrm{EM}},
  \end{equation}
  where $\tensor{{\Pi}}{_{ij}^{lm}} = \Pi^{i}_{l} \Pi^{j}_{m} - \frac{1}{2} \Pi_{ij} \Pi^{lm}$ is the traceless and transverse projector, with $\Pi_{ij} = \delta_{ij} - \partial_{i} \partial_{j} / \Delta$. $T_{lm}^{\mathrm{EM}}$ denotes the spatial component of the energy-momentum tensor of the gauge field, given by $T_{ij}^{\mathrm{EM}} = -a^{2}(E_{i}E_{j} + B_{i} B_{j}) + (\cdots) \, \delta_{ij}$. The $\delta_{ij}$ term vanishes upon contraction with the projection operator and thus does not contribute to the GW source. We then decompose $h_{ij}$ in Fourier space as
  \begin{align}
  \notag
\hat{h}_{ij}(\tau, \bm{x})
  =&\int \frac{\mathrm{d}^{3} \bm{k}}{(2 \pi)^{3 / 2}}
  \hat{h}_{ij}(\tau, \bm{k}) \\
    \notag
    =& \int \frac{\mathrm{d}^{3} \bm{k}}{(2 \pi)^{3 / 2}}
    \sum_{\lambda = +, -}
  (
   \sqrt{2} \epsilon_{i}^{\lambda}(\hat{k}) \epsilon_{j}^{\lambda}(\hat{k})
   \hat{a}^{\lambda}(\bm{k}) \\
   & \times h^{\lambda}(\tau, \bm{k}) \MaE^{ \MaI \bm{k} \cdot \bm{x} }
   + \text{h.c.}),
  \label{eq: hq_Fourier}
  \end{align}
  where $\epsilon_{i}^{\lambda}(\hat{k})$ is the circular polarization vector basis defined earlier. We introduce the projection operator $\Pi^{\pm}_{lm}(\bm{k}) \equiv \epsilon^{\mp}_{l}(\hat{k}) \epsilon^{\mp}_{m}(\hat{k})$, such that $h^{\pm}(\bm{k}) = \Pi_{ij}^{\pm}(\bm{k}) h_{ij}(\bm{k})$. Equation \eqref{eq: hij_eom} can be decomposed into a vacuum part and a sourced part. The vacuum part satisfies the standard Bunch–Davies initial condition and has no source term, while the sourced part satisfies homogeneous initial conditions and can be solved using the Green’s function method. From this point forward, we focus solely on the sourced component and denote it by $h^{\lambda}$. The solution to the sourced part of Eq. \eqref{eq: hij_eom} in Fourier space is given by
  \begin{align}
  \notag
\hat{h}^{\pm} (\tau, \bm{k})
  & = - \dfrac{2 H^{2}}{M^{2}_{\mathrm{pl}}}
  \int_{-\infty}^{\tau} \mathrm{d}\tau' G_{k}(\tau, \tau') \tau'^{2}
  \int \dfrac{\mathrm{d}^{3} \bm{q}}{(2 \pi)^{3 / 2}}
  \Pi_{lm}^{\pm}(\bm{k}) \\
    \notag
    & \times \Bigl[
      \tilde{A}'_{l}( \bm{q}, \tau' ) \hat{A}_{m}'( \bm{k} - \bm{q}, \tau') \\
        \label{eq: sourced_tensor}
        & - \varepsilon_{lab} q_{a} \hat{A}_{b}(\bm{q}, \tau')
        \varepsilon_{mcd}( k_{c} - q_{c}) \hat{A}_{d} (\bm{k} - \bm{q}, \tau')
        \Bigr],
        \end{align}
        where the Green's function is given by
        \begin{align}
        \notag
        G_{k}(\tau, \tau') =& \dfrac{1}{k^{3} \tau'^{2}}
        \bigl[
          (1 + k^{2} \tau \tau') \sin(k (\tau - \tau')) \\
            \label{eq: gw_green_func}
            & - k (\tau - \tau') \cos(k (\tau - \tau')) \bigr] \Theta(\tau - \tau').
            \end{align}
            The power spectrum of the tensor modes is defined by
            \begin{align}
            & \braket{h_{ij}(k, \tau) h_{ij}(k', \tau)}
            \equiv \dfrac{2 \pi^{2}}{k^{3}} \mathcal{P}_{h}(k, \tau) \delta(\bm{k} + \bm{k}'), \\
              & \braket{h^{\pm}(k, \tau) h^{\pm}(k', \tau)}
              \equiv \dfrac{2 \pi^{2}}{k^{3}} \mathcal{P}^{\pm}_{h}(k, \tau) \delta(\bm{k} + \bm{k}'),
              \end{align}
              which immediately gives $\mathcal{P}_{h} = \mathcal{P}^{+}_{h} + \mathcal{P}^{-}_{h}$.
Now we can evaluate the GW power spectra by applying Wick’s theorem at the end of the inflation as
\begin{widetext}
\begin{align}
  \notag
  \mathcal{P}_{h}^{\lambda}
  =& \frac{H^{4} k^{3}}{\pi^{4} M_{\mathrm{pl}}^{4}}
  \int_{0}^{\infty} q^{2} \mathrm{d} q
  \int_{-1}^{1} \mathrm{d} u
  \Biggr[
    \left| \epsilon^{\lambda}_{i}(\bm{k})
      \epsilon^{-}_{i}(\bm{-q}) \right|^{2}
    \left| \epsilon^{\lambda}_{j}(\bm{k})
      \epsilon^{-}_{j}(\bm{q} - \bm{k}) \right|^{2} \\
    \label{eq: sourced_tensor_spectrum}
    & \times \biggl| \int_{-\infty}^{0} \mathrm{d} \tau' \tau'^{2}
    G_{k}(\tau_{\mathrm{end}}, \tau')
    \biggl(
      A^{'-}(\tau', \bm{q}) A^{'-}(\tau', \bm{k}-\bm{q})
      + q |\bm{k} - \bm{q}|
      A^{-}(\tau', \bm{q})
      A^{-}(\tau', \bm{k} - \bm{q}) \biggr) \biggr|^{2} \Biggr],
\end{align}
\end{widetext}
where $u \equiv \cos \theta$, $\tau_{\mathrm{end}}$ denotes the time at the end of inflation, and we only retain the enhanced mode $A^{-}(\bm{k})$. The two norms of the polarization vector can be computed via the following property,
\begin{equation}
  \left| \epsilon_{i}^{\lambda}(\hat{p})
    \epsilon_{i}^{\lambda'}(\hat{q}) \right|^{2}
  = \left( \frac{1 - \lambda \lambda' \hat{p} \cdot \hat{q}}{2} \right)^{2}
\end{equation}
Then one can compute today’s energy density of the stochastic GW background, which is related to the GW power spectrum at the end of inflation by \cite{Caprini:2018mtu}
\begin{equation}
  \Omega_{\mathrm{gw},0}h^2 = \frac{\Omega_{r, 0} h^{2}}{24}
  ( \mathcal{P}^{+}_{h}(\tau_{\mathrm{end}},k) + \mathcal{P}^{-}_{h}(\tau_{\mathrm{end}},k) ).
\end{equation}
 Here, $\Omega_{\mathrm{r}, 0} \simeq 4.15 \times 10^{-5} / h^{2}$denotes the current fraction of the radiation energy density to the total energy density of the Universe.

\begin{figure}[tbp]
  \includegraphics[width=.48\textwidth]{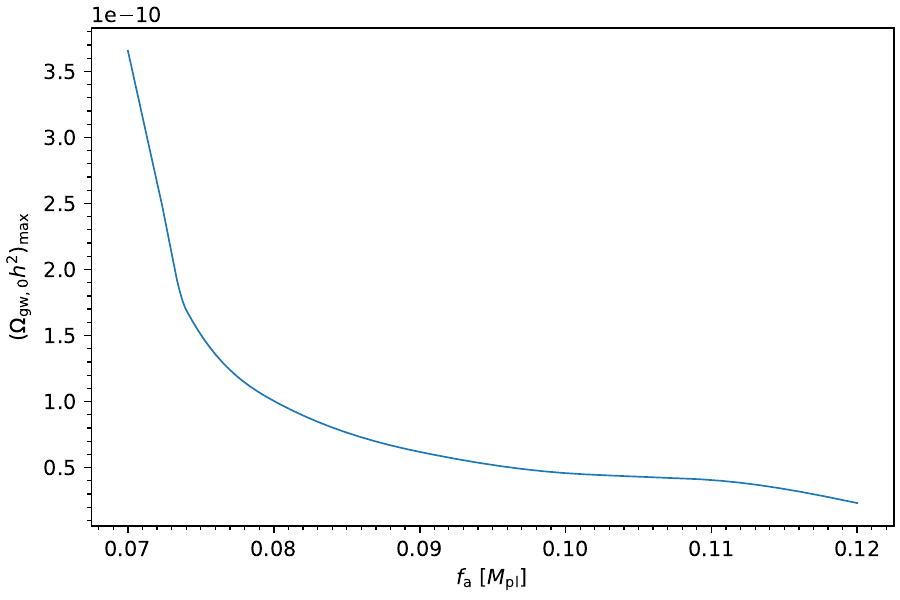}
  \caption{ Maximum value of the GW spectrum $\Omega_{\mathrm{gw},0}h^{2}$ as a function of $f_{\rm a}$. For each value of $f_{\mathrm{a}}$, we scan the parameter space of $\alpha$ to determine the maximum spectrum. Since $\chi$ initially follows an attractor solution, the result is not sensitive to the initial value of $\chi$. }
  \label{fig: gw}
\end{figure}

 Since GWs are sourced by the gauge quanta, the dependence of their energy spectrum on the model parameters is similar to that of the gauge field. That is, to obtain a larger energy spectrum, one needs to decrease $f_{\rm a}$ and choose an appropriate value of $\alpha$. If $\alpha$ is too small, it cannot generate a sufficient amount of gauge quanta, while an excessively large $\alpha$ will suppress their production.
In principle, $\alpha$ is a model-dependent parameter, and its value can range from $\mathcal{O}(1)$ to $\mathcal{O}(100)$ \cite{Anber:2009ua}. Additionally, the condition $f_{\rm a} < 1$ should be satisfied from the perspective of string theory or effective field theory \cite{Barnaby:2011pe}. Figure \ref{fig: gw} shows the maximum value of the GW spectrum for varying $f_{\rm a}$. For each value of $f_{\rm a}$, we scan the parameter space of $\alpha$ to identify the maximum GW signal. The parameters used in the numerical computation safely satisfy the aforementioned constraints. Since $\chi$ initially follows an attractor solution, our results are not sensitive to its initial value. The results indicate that a smaller $f_{\rm a}$ leads to a larger $\Omega_{\mathrm{gw},0}$, although the order of magnitude remains approximately unchanged, with $\Omega_{\mathrm{gw},0} h^{2} \sim 10^{-10}$.

\section{Initial condition of the axion}
\label{sec: ini_cond}

\begin{figure}[tbp]
  \includegraphics[width=.48\textwidth]{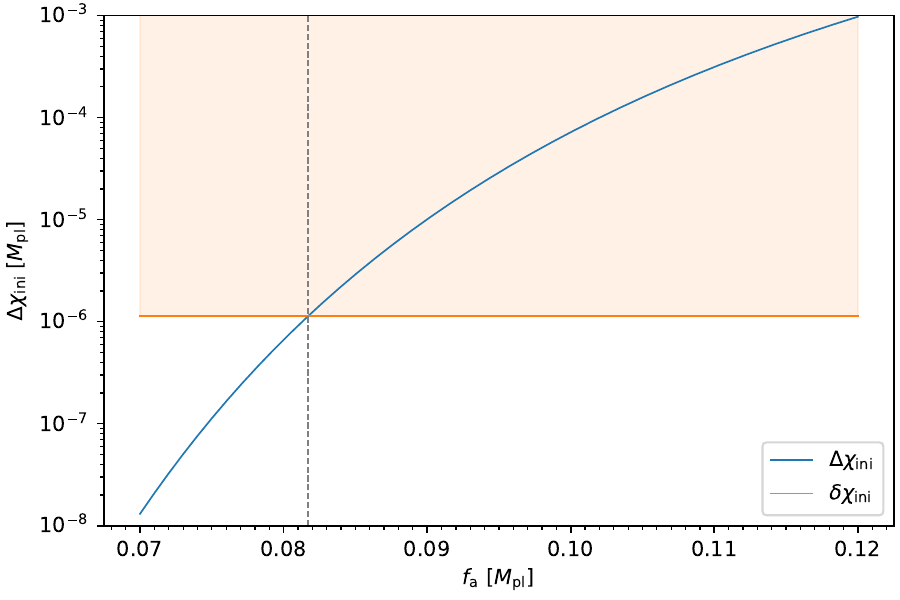}
  \caption{
  Constraint on the parameter $f_{\rm a}$ for GW spectrum peaking at nHz region. The corresponding {\it e}-folding number is computed as $ N_{*} = N_{\mathrm{p}} - 15.37$. The blue curve shows the variation of $\Delta \chi_{\mathrm{ini}}$ with respect to the parameter $f_{\rm a}$, plotted based on the LHS of Eq. \eqref{eq: f_constrain}. The orange curve represents the uncertainty in the initial value of the axion due to the vacuum fluctuations, $\delta \chi_{\mathrm{ini}}$, plotted according to the RHS of Eq. \eqref{eq: f_constrain}. The shadow area which lies above the orange curve is physically allowed region. The black dashed line corresponds to $f_{\rm a} = 0.0817$ at which $\delta \chi_{\mathrm{ini}} = \Delta \chi_{\mathrm{ini}}$.
  }
  \label{fig: initial_nhz}
\end{figure}

\begin{figure}[tbp]
  \includegraphics[width=.48\textwidth]{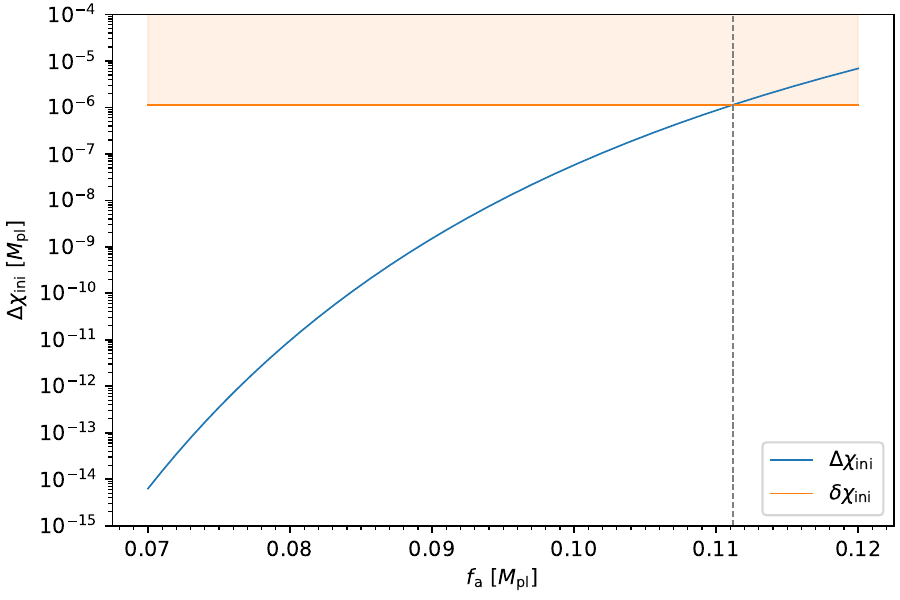}
  \caption{
  Constraint on the parameter $f_{\rm a}$ for GW spectrum peaking at mHz region. The corresponding {\it e}-folding number is computed as $ N_{*} = N_{\mathrm{p}} - 29.19$. The blue curve shows the variation of $\Delta \chi_{\mathrm{ini}}$ with respect to the parameter $f_{\rm a}$, plotted based on the LHS of Eq. \eqref{eq: f_constrain}. The orange curve represents the uncertainty in the initial value of the axion due to the vacuum fluctuations, $\delta \chi_{\mathrm{ini}}$, plotted according to the RHS of Eq. \eqref{eq: f_constrain}. The shadow area which lies above the orange curve is physically allowed region. The black dashed line corresponds to $f_{\rm a} = 0.111$ at which $\delta \chi_{\mathrm{ini}} = \Delta \chi_{\mathrm{ini}}$.
  }
  \label{fig: initial_mhz}
\end{figure}
An observable GW signal must satisfy two basic requirements: (i) its maximum energy spectrum must reach the sensitivity of the observational instrument, and (ii) the corresponding peak frequency must lie within the instrument’s most sensitive frequency range. In the previous section, we discussed the maximum GW energy spectrum, which addresses the first requirement. In this section, we focus on how the initial value of the axion influences the peak frequency, which is relevant to the second requirement.

In Section \ref{sec: dynamics}, we discussed the three stages of the system’s dynamics. Initially, the axion slowly rolls from the peak of the potential. It then undergoes a fast-roll phase, during which gauge quanta are exponentially produced; these quanta also source GWs. Finally, the production of gauge quanta becomes strong enough to slow down the axion field, and the system enters the strong backreaction regime. Typically, the total {\it{e}}-folding number during the slow-roll phase spans several dozen, while the fast-roll and strong backreaction phases each last less than a few {\it{e}}-foldings. The e-folding number $N$ at which a mode $k$ exits the horizon and the corresponding frequency of that mode $f = {k}/{(2\pi)}$ are related by \cite{Ozsoy:2020ccy}
\begin{align}
  \notag
  N_{\mathrm{p}} - N
  & = 41.7 - \ln \left( \frac{k_{\mathrm{p}}}{0.05 \mathrm{Mpc}^{-1}} \right) \\
  \label{eq: efold-f-1}
  & \mathrel{\phantom{=}}
  + \ln \left( \frac{f}{100 \mathrm{Hz}} \right)
  - \ln \left( \frac{H_N}{H_\mathrm{p}} \right),
\end{align}
where the subscript $\rm{p}$ denotes the CMB pivot scale, and $H_{N}$ represents the Hubble parameter at {\it{e}}-folding number $N$. Typically, $N_{\mathrm{p}} \sim 60$ and $k_{\mathrm{p}} \sim 0.05, \mathrm{Mpc}^{-1}$. Due to the slow-roll nature of the background, we have $H_{N} \sim H_{\mathrm{p}}$. Therefore, the above equation simplifies to
\begin{equation}
  \label{eq: efold-f-2}
  19.3 - N \simeq \ln \left(\frac{f}{100 \mathrm{Hz}}\right).
\end{equation}
Equation \eqref{eq: efold-f-1} also implies $\Delta N = -\Delta \ln(f)$.

The GW signal with momentum $k$ primarily depends on the production of gauge quanta with similar momentum (a detailed analysis can be found in our previous work \cite{He:2025ieo}). Since the production of $A^{-}(k)$ mainly occurs near the horizon-crossing time, denoted as $N(k)$, the corresponding generation of $h_{ij}(k)$ also takes place at that time. Moreover, the peak of particle production coincides with the onset of strong backreaction, which closely aligns with the moment when the axion enters its fast-roll phase. Therefore, the peak position of the GW spectrum is mainly determined by the duration of the slow-roll stage: a longer slow-roll phase results in a peak at higher frequency.

Due to quantum fluctuations, there exists an intrinsic uncertainty in the initial value of the axion field. Therefore, the initial value cannot be arbitrarily close to the peak of the potential, which in turn constrains the possible position of the GW spectrum peak. From the potential in eq. \eqref{eq: Uchi}, we see that a smaller $f_{\rm a}$ leads to a sharper potential peak, corresponding to a shorter slow-roll duration. In other words, although a smaller $f_{\rm a}$ can slightly enhance the maximum GW amplitude, it also restricts the allowed frequency range for the GW peak. As we will show shortly, for a given peak frequency, there exists a minimum value of $f_{\mathrm{a}}$ corresponding to it.

Using the slow-roll approximation and noting that the backreaction is negligible at this stage, the EoM of the axion field given in Eq. \eqref{eq: chi_eof} reduces to
\begin{equation}
  \label{eq: chi_eof_slowroll}
  3H \dot{\chi} + U_{,\chi} = 0.
\end{equation}
Equation \eqref{eq: chi_eof_slowroll} admits an analytic solution \cite{Namba:2015gja} (note that our convention is different from that of \cite{Namba:2015gja}, as we define $\mathrm{d}N = -H\mathrm{d}t$):
\begin{equation}
  \label{eq: chi_eof_slowroll_sol}
  \chi(N) = 2f_{\rm a} \left( \pi
  - \arctan \left( e^{\frac{\Lambda^{4}}{6H^{2}f_{\rm a}^{2}} (N_{*} - N)} \right)
  \right),
\end{equation}
Without loss of generality, we assume that the axion initially rolls from the left side of the peak position at $\chi / f_{\rm a} = 2 \pi$ with a negative initial velocity, $\dot{\chi}_{\mathrm{ini}} < 0$. Here, $N_{*}$ denotes the time when axion reaches the rapidest point $\chi / f_{\rm a} = \frac{2}{3}\pi$ of the potential, which can be considered as the typical onset time of the fast-roll phase.
By setting $N = N_{\mathrm{ini}}$, we can determine the distance between the initial value $\chi_{\mathrm{ini}}$ and the position of the peak of the potential
\begin{equation}
  \label{eq: delta_chi_ini}
  \Delta \chi_{\mathrm{ini}} \equiv 2 \MaPI f_{\mathrm{a}} - \chi(N_{\mathrm{ini}})
  = 2 f_{\mathrm{a}}
  \mathrm{e}^{-\frac{\Lambda^{4}}{6H^{2}f_{\rm a}^{2}} (N_{\mathrm{ini}} - N_{*})}
\end{equation}
On the other hand, the uncertainty arising from quantum fluctuations can be estimated by
\begin{align}
  \label{eq: delta_chi_qmvac}
  \delta \chi_{\mathrm{ini}} \sim \sqrt{\mathcal{P}_{\delta \chi}(k)} = \frac{H}{2\pi},
\end{align}
where we used $\mathcal{P}_{\delta \chi} = \left( {H}/{(2 \pi)} \right)^{2}$.

In this way, for a given frequency $f_{*}$, one can compute the corresponding $N_{*}$ and $\Delta \chi_{\mathrm{ini}}$ from Eq. \eqref{eq: efold-f-2} and  Eq. \eqref{eq: delta_chi_ini}, respectively.
As mentioned earlier, due to quantum fluctuations, the initial value cannot be arbitrarily close to the peak of the potential, which requires that $\Delta \chi_{\mathrm{ini}} > \delta \chi_{\mathrm{ini}}$. This condition, in turn, imposes a constraint on the minimal value of the parameter $f_{\mathrm{a}}$:
\begin{equation}
  \label{eq: f_constrain}
  2 f_{\mathrm{a}}
  \mathrm{e}^{-\frac{\Lambda^{4}}{6H^{2}f_{\rm a}^{2}} (N_{\mathrm{ini}} - N_{*})}
  > \frac{H}{2 \pi}.
\end{equation}
Figures \ref{fig: initial_nhz} and \ref{fig: initial_mhz} present the resulting constraints on $f_{\rm a}$ based on the previous analysis, where the peak frequencies lie in the nHz and mHz ranges, respectively. The blue curve is plotted based on the LHS of Eq. \eqref{eq: f_constrain}, while the orange curve is plotted based on the RHS of Eq. \eqref{eq: f_constrain}. The physically allowed area, where $\Delta \chi_{\mathrm{ini}} < \delta \chi_{\mathrm{ini}}$, is indicated by the shadow area. Our results indicate that for GWs peaking in the nHz region, the allowed values of $f_{\rm a}$ satisfy $f_{\rm a} > 0.0817$, and for the mHz region, $f_{\rm a} > 0.111$. Combining these with Fig. \ref{fig: gw}, we find that the maximum allowed $\Omega_{\mathrm{GW},0}$ is approximately $\Omega_{\mathrm{GW},0} \sim 10^{-10}$ in both cases. While such an amplitude is insufficient to account for the recent pulsar timing array (PTA) data, it remains potentially detectable by future mHz experiments such as LISA and Taiji.


\section{Conclusions}
\label{sec: conclusion}

In this work, we investigate the production of GWs from a gauge field coupled to a spectator axion field through the Chern-Simons interaction during inflation, with a particular focus on the strong backreaction regime. The axion evolves in a cosine potential, while inflation is driven by a background scalar field consistent with current CMB observations. The system's dynamics can be divided into three stages. Initially, the axion slowly rolls from the peak of the potential; this stage typically lasts for several dozen {\it e}-foldings. Then, the axion enters a fast-roll phase, during which gauge quanta are exponentially produced and subsequently source GWs. As the production of gauge quanta becomes substantial, the backreaction grows strong enough to decelerate the axion, suppressing further particle production.The axion then accelerates again. This repeating process results in an oscillating axion speed and a GW spectrum with multiple peaks.
We omit the axion inhomogeneous term throughout this work. This term has been thoroughly investigated using lattice simulations near the end of inflation in previous studies \cite{Caravano:2022epk, Caravano:2024xsb, Sharma:2024nfu, Figueroa:2024rkr, Lizarraga:2025aiw, Jamieson:2025ngu, Caravano:2021bfn, Figueroa:2023oxc}. These works show that the inhomogeneous term can suppress oscillatory behavior when the energy density of the gauge field becomes comparable to the axion potential. In contrast, our analysis focuses on the fast-roll stage, where the energy density of the gauge field is always much smaller than the axion potential. Therefore, the axion inhomogeneous term is negligible in our case.

Since the maximum value of the GW energy spectrum is mainly influenced by the dynamics following the slow-roll phase of the axion—and the slow-roll stage is an attractor solution—this result is not sensitive to the initial value of the axion field. As expected, we find that a smaller value of $f_a$ leads to a larger maximum GW amplitude. However, this dependence is relatively weak, and within the parameter range of interest, it does not significantly alter the order of magnitude. Across the parameter space explored in this study, we find that the maximum value of the GW energy spectrum is approximately $\Omega_{\mathrm{gw},0} h^{2} = 10^{-10}$.

On the other hand, due to quantum fluctuations, the initial value of the axion field inherently carries an intrinsic uncertainty, which prevents it from being arbitrarily close to the peak of the potential and, in turn, constrains the allowed duration of the slow-roll stage. Since $f_{\rm a}$ controls the sharpness of the potential, a smaller $f_{\mathrm{a}}$ corresponds to a shorter slow-roll phase. For a fixed peak frequency, by requiring that the distance between the initial value and the potential peak, $\Delta \chi_{\mathrm{ini}}$, be larger than the quantum fluctuation, we derive a constraint on $f_{\rm a}$, as given by Eq. \eqref{eq: f_constrain}.

We then apply this result to determine the lowest possible value of $f_{\rm a}$ for GW peaks appearing in the nHz and mHz frequency ranges. The results are shown in Figs. \ref{fig: initial_nhz} and \ref{fig: initial_mhz}. We find that, despite the differences in frequency range, the weak dependence of $\Omega_{\mathrm{gw}, 0} h^{2}$ on the parameter $f_{\rm a}$ leads to a similar maximum allowed GW energy spectrum in both cases, namely $\Omega_{\mathrm{gw},0} h^{2} = 10^{-10}$. Consequently, given this amplitude, GW signals produced in this model with the chosen parameters are unlikely to account for the recent PTA results, but they may still be detectable by future experiments targeting the mHz frequency range, such as Taiji and LISA.
It is possible to obtain stronger GWs and potentially reach the PTA range with a larger $\Lambda^{4}$, even after accounting for the constraint from quantum fluctuations. However, if $\Lambda^{4}$ becomes too large, for example $\Lambda^{4} = 0.1 \, V_{0}$, it begins to affect the total number of $e$-folds and shift the peak position of the GW spectrum. In this case, more fine-tuning is required to achieve the desired peak position, making the model less natural.


\begin{acknowledgments}
This work is supported in part by the National Key Research and Development Program of China Grant No. 2020YFC2201501, in part by the National Natural Science Foundation of China under Grant No. 12305057, No. 12475067, and No. 12235019.

\end{acknowledgments}

\bibliography{reference} 

\begin{thebibliography}{119}%
\makeatletter
\providecommand \@ifxundefined [1]{%
 \@ifx{#1\undefined}
}%
\providecommand \@ifnum [1]{%
 \ifnum #1\expandafter \@firstoftwo
 \else \expandafter \@secondoftwo
 \fi
}%
\providecommand \@ifx [1]{%
 \ifx #1\expandafter \@firstoftwo
 \else \expandafter \@secondoftwo
 \fi
}%
\providecommand \natexlab [1]{#1}%
\providecommand \enquote  [1]{``#1''}%
\providecommand \bibnamefont  [1]{#1}%
\providecommand \bibfnamefont [1]{#1}%
\providecommand \citenamefont [1]{#1}%
\providecommand \href@noop [0]{\@secondoftwo}%
\providecommand \href [0]{\begingroup \@sanitize@url \@href}%
\providecommand \@href[1]{\@@startlink{#1}\@@href}%
\providecommand \@@href[1]{\endgroup#1\@@endlink}%
\providecommand \@sanitize@url [0]{\catcode `\\12\catcode `\$12\catcode `\&12\catcode `\#12\catcode `\^12\catcode `\_12\catcode `\%12\relax}%
\providecommand \@@startlink[1]{}%
\providecommand \@@endlink[0]{}%
\providecommand \url  [0]{\begingroup\@sanitize@url \@url }%
\providecommand \@url [1]{\endgroup\@href {#1}{\urlprefix }}%
\providecommand \urlprefix  [0]{URL }%
\providecommand \Eprint [0]{\href }%
\providecommand \doibase [0]{https://doi.org/}%
\providecommand \selectlanguage [0]{\@gobble}%
\providecommand \bibinfo  [0]{\@secondoftwo}%
\providecommand \bibfield  [0]{\@secondoftwo}%
\providecommand \translation [1]{[#1]}%
\providecommand \BibitemOpen [0]{}%
\providecommand \bibitemStop [0]{}%
\providecommand \bibitemNoStop [0]{.\EOS\space}%
\providecommand \EOS [0]{\spacefactor3000\relax}%
\providecommand \BibitemShut  [1]{\csname bibitem#1\endcsname}%
\let\auto@bib@innerbib\@empty
\bibitem [{\citenamefont {Guth}(1981)}]{Guth:1980zm}%
  \BibitemOpen
  \bibfield  {author} {\bibinfo {author} {\bibfnamefont {A.~H.}\ \bibnamefont {Guth}},\ }\bibfield  {title} {\bibinfo {title} {{The Inflationary Universe: A Possible Solution to the Horizon and Flatness Problems}},\ }\href {https://doi.org/10.1103/PhysRevD.23.347} {\bibfield  {journal} {\bibinfo  {journal} {Phys. Rev. D}\ }\textbf {\bibinfo {volume} {23}},\ \bibinfo {pages} {347} (\bibinfo {year} {1981})}\BibitemShut {NoStop}%
\bibitem [{\citenamefont {Sato}(1981)}]{Sato:1980yn}%
  \BibitemOpen
  \bibfield  {author} {\bibinfo {author} {\bibfnamefont {K.}~\bibnamefont {Sato}},\ }\bibfield  {title} {\bibinfo {title} {{First Order Phase Transition of a Vacuum and Expansion of the Universe}},\ }\href@noop {} {\bibfield  {journal} {\bibinfo  {journal} {Mon. Not. Roy. Astron. Soc.}\ }\textbf {\bibinfo {volume} {195}},\ \bibinfo {pages} {467} (\bibinfo {year} {1981})}\BibitemShut {NoStop}%
\bibitem [{\citenamefont {Linde}(1982)}]{Linde:1981mu}%
  \BibitemOpen
  \bibfield  {author} {\bibinfo {author} {\bibfnamefont {A.~D.}\ \bibnamefont {Linde}},\ }\bibfield  {title} {\bibinfo {title} {{A New Inflationary Universe Scenario: A Possible Solution of the Horizon, Flatness, Homogeneity, Isotropy and Primordial Monopole Problems}},\ }\href {https://doi.org/10.1016/0370-2693(82)91219-9} {\bibfield  {journal} {\bibinfo  {journal} {Phys. Lett. B}\ }\textbf {\bibinfo {volume} {108}},\ \bibinfo {pages} {389} (\bibinfo {year} {1982})}\BibitemShut {NoStop}%
\bibitem [{\citenamefont {Albrecht}\ and\ \citenamefont {Steinhardt}(1982)}]{Albrecht:1982wi}%
  \BibitemOpen
  \bibfield  {author} {\bibinfo {author} {\bibfnamefont {A.}~\bibnamefont {Albrecht}}\ and\ \bibinfo {author} {\bibfnamefont {P.~J.}\ \bibnamefont {Steinhardt}},\ }\bibfield  {title} {\bibinfo {title} {{Cosmology for Grand Unified Theories with Radiatively Induced Symmetry Breaking}},\ }\href {https://doi.org/10.1103/PhysRevLett.48.1220} {\bibfield  {journal} {\bibinfo  {journal} {Phys. Rev. Lett.}\ }\textbf {\bibinfo {volume} {48}},\ \bibinfo {pages} {1220} (\bibinfo {year} {1982})}\BibitemShut {NoStop}%
\bibitem [{\citenamefont {Starobinsky}(1980)}]{Starobinsky:1980te}%
  \BibitemOpen
  \bibfield  {author} {\bibinfo {author} {\bibfnamefont {A.~A.}\ \bibnamefont {Starobinsky}},\ }\bibfield  {title} {\bibinfo {title} {{A New Type of Isotropic Cosmological Models Without Singularity}},\ }\href {https://doi.org/10.1016/0370-2693(80)90670-X} {\bibfield  {journal} {\bibinfo  {journal} {Phys. Lett. B}\ }\textbf {\bibinfo {volume} {91}},\ \bibinfo {pages} {99} (\bibinfo {year} {1980})}\BibitemShut {NoStop}%
\bibitem [{\citenamefont {Starobinsky}(1979)}]{Starobinsky:1979ty}%
  \BibitemOpen
  \bibfield  {author} {\bibinfo {author} {\bibfnamefont {A.~A.}\ \bibnamefont {Starobinsky}},\ }\bibfield  {title} {\bibinfo {title} {{Spectrum of relict gravitational radiation and the early state of the universe}},\ }\href@noop {} {\bibfield  {journal} {\bibinfo  {journal} {JETP Lett.}\ }\textbf {\bibinfo {volume} {30}},\ \bibinfo {pages} {682} (\bibinfo {year} {1979})}\BibitemShut {NoStop}%
\bibitem [{\citenamefont {Mukhanov}\ and\ \citenamefont {Chibisov}(1981)}]{Mukhanov:1981xt}%
  \BibitemOpen
  \bibfield  {author} {\bibinfo {author} {\bibfnamefont {V.~F.}\ \bibnamefont {Mukhanov}}\ and\ \bibinfo {author} {\bibfnamefont {G.~V.}\ \bibnamefont {Chibisov}},\ }\bibfield  {title} {\bibinfo {title} {{Quantum Fluctuations and a Nonsingular Universe}},\ }\href@noop {} {\bibfield  {journal} {\bibinfo  {journal} {JETP Lett.}\ }\textbf {\bibinfo {volume} {33}},\ \bibinfo {pages} {532} (\bibinfo {year} {1981})}\BibitemShut {NoStop}%
\bibitem [{\citenamefont {Hawking}(1982)}]{Hawking:1982cz}%
  \BibitemOpen
  \bibfield  {author} {\bibinfo {author} {\bibfnamefont {S.~W.}\ \bibnamefont {Hawking}},\ }\bibfield  {title} {\bibinfo {title} {{The Development of Irregularities in a Single Bubble Inflationary Universe}},\ }\href {https://doi.org/10.1016/0370-2693(82)90373-2} {\bibfield  {journal} {\bibinfo  {journal} {Phys. Lett. B}\ }\textbf {\bibinfo {volume} {115}},\ \bibinfo {pages} {295} (\bibinfo {year} {1982})}\BibitemShut {NoStop}%
\bibitem [{\citenamefont {Guth}\ and\ \citenamefont {Pi}(1982)}]{Guth:1982ec}%
  \BibitemOpen
  \bibfield  {author} {\bibinfo {author} {\bibfnamefont {A.~H.}\ \bibnamefont {Guth}}\ and\ \bibinfo {author} {\bibfnamefont {S.~Y.}\ \bibnamefont {Pi}},\ }\bibfield  {title} {\bibinfo {title} {{Fluctuations in the New Inflationary Universe}},\ }\href {https://doi.org/10.1103/PhysRevLett.49.1110} {\bibfield  {journal} {\bibinfo  {journal} {Phys. Rev. Lett.}\ }\textbf {\bibinfo {volume} {49}},\ \bibinfo {pages} {1110} (\bibinfo {year} {1982})}\BibitemShut {NoStop}%
\bibitem [{\citenamefont {Starobinsky}(1982)}]{Starobinsky:1982ee}%
  \BibitemOpen
  \bibfield  {author} {\bibinfo {author} {\bibfnamefont {A.~A.}\ \bibnamefont {Starobinsky}},\ }\bibfield  {title} {\bibinfo {title} {{Dynamics of Phase Transition in the New Inflationary Universe Scenario and Generation of Perturbations}},\ }\href {https://doi.org/10.1016/0370-2693(82)90541-X} {\bibfield  {journal} {\bibinfo  {journal} {Phys. Lett. B}\ }\textbf {\bibinfo {volume} {117}},\ \bibinfo {pages} {175} (\bibinfo {year} {1982})}\BibitemShut {NoStop}%
\bibitem [{\citenamefont {Abbott}\ and\ \citenamefont {Wise}(1984)}]{Abbott:1984fp}%
  \BibitemOpen
  \bibfield  {author} {\bibinfo {author} {\bibfnamefont {L.~F.}\ \bibnamefont {Abbott}}\ and\ \bibinfo {author} {\bibfnamefont {M.~B.}\ \bibnamefont {Wise}},\ }\bibfield  {title} {\bibinfo {title} {{Constraints on Generalized Inflationary Cosmologies}},\ }\href {https://doi.org/10.1016/0550-3213(84)90329-8} {\bibfield  {journal} {\bibinfo  {journal} {Nucl. Phys. B}\ }\textbf {\bibinfo {volume} {244}},\ \bibinfo {pages} {541} (\bibinfo {year} {1984})}\BibitemShut {NoStop}%
\bibitem [{\citenamefont {Lyth}(1984)}]{Lyth:1984yz}%
  \BibitemOpen
  \bibfield  {author} {\bibinfo {author} {\bibfnamefont {D.~H.}\ \bibnamefont {Lyth}},\ }\bibfield  {title} {\bibinfo {title} {{A Bound on Inflationary Energy Density From the Isotropy of the Microwave Background}},\ }\href {https://doi.org/10.1016/0370-2693(84)91391-1} {\bibfield  {journal} {\bibinfo  {journal} {Phys. Lett. B}\ }\textbf {\bibinfo {volume} {147}},\ \bibinfo {pages} {403} (\bibinfo {year} {1984})},\ \bibinfo {note} {[Erratum: Phys.Lett.B 150, 465 (1985)]}\BibitemShut {NoStop}%
\bibitem [{\citenamefont {Lyth}(1997)}]{Lyth:1996im}%
  \BibitemOpen
  \bibfield  {author} {\bibinfo {author} {\bibfnamefont {D.~H.}\ \bibnamefont {Lyth}},\ }\bibfield  {title} {\bibinfo {title} {{What would we learn by detecting a gravitational wave signal in the cosmic microwave background anisotropy?}},\ }\href {https://doi.org/10.1103/PhysRevLett.78.1861} {\bibfield  {journal} {\bibinfo  {journal} {Phys. Rev. Lett.}\ }\textbf {\bibinfo {volume} {78}},\ \bibinfo {pages} {1861} (\bibinfo {year} {1997})},\ \Eprint {https://arxiv.org/abs/hep-ph/9606387} {arXiv:hep-ph/9606387} \BibitemShut {NoStop}%
\bibitem [{\citenamefont {Boubekeur}(2013)}]{Boubekeur:2012xn}%
  \BibitemOpen
  \bibfield  {author} {\bibinfo {author} {\bibfnamefont {L.}~\bibnamefont {Boubekeur}},\ }\bibfield  {title} {\bibinfo {title} {{Theoretical bounds on the tensor-to-scalar ratio in the cosmic microwave background}},\ }\href {https://doi.org/10.1103/PhysRevD.87.061301} {\bibfield  {journal} {\bibinfo  {journal} {Phys. Rev. D}\ }\textbf {\bibinfo {volume} {87}},\ \bibinfo {pages} {061301} (\bibinfo {year} {2013})},\ \Eprint {https://arxiv.org/abs/1208.0210} {arXiv:1208.0210 [astro-ph.CO]} \BibitemShut {NoStop}%
\bibitem [{\citenamefont {Baumann}\ and\ \citenamefont {McAllister}(2007)}]{Baumann:2006cd}%
  \BibitemOpen
  \bibfield  {author} {\bibinfo {author} {\bibfnamefont {D.}~\bibnamefont {Baumann}}\ and\ \bibinfo {author} {\bibfnamefont {L.}~\bibnamefont {McAllister}},\ }\bibfield  {title} {\bibinfo {title} {{A Microscopic Limit on Gravitational Waves from D-brane Inflation}},\ }\href {https://doi.org/10.1103/PhysRevD.75.123508} {\bibfield  {journal} {\bibinfo  {journal} {Phys. Rev. D}\ }\textbf {\bibinfo {volume} {75}},\ \bibinfo {pages} {123508} (\bibinfo {year} {2007})},\ \Eprint {https://arxiv.org/abs/hep-th/0610285} {arXiv:hep-th/0610285} \BibitemShut {NoStop}%
\bibitem [{\citenamefont {Akrami}\ \emph {et~al.}(2020{\natexlab{a}})\citenamefont {Akrami} \emph {et~al.}}]{Planck:2018jri}%
  \BibitemOpen
  \bibfield  {author} {\bibinfo {author} {\bibfnamefont {Y.}~\bibnamefont {Akrami}} \emph {et~al.} (\bibinfo {collaboration} {Planck}),\ }\bibfield  {title} {\bibinfo {title} {{Planck 2018 results. X. Constraints on inflation}},\ }\href {https://doi.org/10.1051/0004-6361/201833887} {\bibfield  {journal} {\bibinfo  {journal} {Astron. Astrophys.}\ }\textbf {\bibinfo {volume} {641}},\ \bibinfo {pages} {A10} (\bibinfo {year} {2020}{\natexlab{a}})},\ \Eprint {https://arxiv.org/abs/1807.06211} {arXiv:1807.06211 [astro-ph.CO]} \BibitemShut {NoStop}%
\bibitem [{\citenamefont {Akrami}\ \emph {et~al.}(2020{\natexlab{b}})\citenamefont {Akrami} \emph {et~al.}}]{Planck:2019kim}%
  \BibitemOpen
  \bibfield  {author} {\bibinfo {author} {\bibfnamefont {Y.}~\bibnamefont {Akrami}} \emph {et~al.} (\bibinfo {collaboration} {Planck}),\ }\bibfield  {title} {\bibinfo {title} {{Planck 2018 results. IX. Constraints on primordial non-Gaussianity}},\ }\href {https://doi.org/10.1051/0004-6361/201935891} {\bibfield  {journal} {\bibinfo  {journal} {Astron. Astrophys.}\ }\textbf {\bibinfo {volume} {641}},\ \bibinfo {pages} {A9} (\bibinfo {year} {2020}{\natexlab{b}})},\ \Eprint {https://arxiv.org/abs/1905.05697} {arXiv:1905.05697 [astro-ph.CO]} \BibitemShut {NoStop}%
\bibitem [{\citenamefont {Tristram}\ \emph {et~al.}(2021)\citenamefont {Tristram} \emph {et~al.}}]{Tristram:2020wbi}%
  \BibitemOpen
  \bibfield  {author} {\bibinfo {author} {\bibfnamefont {M.}~\bibnamefont {Tristram}} \emph {et~al.},\ }\bibfield  {title} {\bibinfo {title} {{Planck constraints on the tensor-to-scalar ratio}},\ }\href {https://doi.org/10.1051/0004-6361/202039585} {\bibfield  {journal} {\bibinfo  {journal} {Astron. Astrophys.}\ }\textbf {\bibinfo {volume} {647}},\ \bibinfo {pages} {A128} (\bibinfo {year} {2021})},\ \Eprint {https://arxiv.org/abs/2010.01139} {arXiv:2010.01139 [astro-ph.CO]} \BibitemShut {NoStop}%
\bibitem [{\citenamefont {Ade}\ \emph {et~al.}(2018)\citenamefont {Ade} \emph {et~al.}}]{BICEP2:2018kqh}%
  \BibitemOpen
  \bibfield  {author} {\bibinfo {author} {\bibfnamefont {P.~A.~R.}\ \bibnamefont {Ade}} \emph {et~al.} (\bibinfo {collaboration} {BICEP2, Keck Array}),\ }\bibfield  {title} {\bibinfo {title} {{BICEP2 / Keck Array x: Constraints on Primordial Gravitational Waves using Planck, WMAP, and New BICEP2/Keck Observations through the 2015 Season}},\ }\href {https://doi.org/10.1103/PhysRevLett.121.221301} {\bibfield  {journal} {\bibinfo  {journal} {Phys. Rev. Lett.}\ }\textbf {\bibinfo {volume} {121}},\ \bibinfo {pages} {221301} (\bibinfo {year} {2018})},\ \Eprint {https://arxiv.org/abs/1810.05216} {arXiv:1810.05216 [astro-ph.CO]} \BibitemShut {NoStop}%
\bibitem [{\citenamefont {Ade}\ \emph {et~al.}(2021)\citenamefont {Ade} \emph {et~al.}}]{BICEP:2021xfz}%
  \BibitemOpen
  \bibfield  {author} {\bibinfo {author} {\bibfnamefont {P.~A.~R.}\ \bibnamefont {Ade}} \emph {et~al.} (\bibinfo {collaboration} {BICEP, Keck}),\ }\bibfield  {title} {\bibinfo {title} {{Improved Constraints on Primordial Gravitational Waves using Planck, WMAP, and BICEP/Keck Observations through the 2018 Observing Season}},\ }\href {https://doi.org/10.1103/PhysRevLett.127.151301} {\bibfield  {journal} {\bibinfo  {journal} {Phys. Rev. Lett.}\ }\textbf {\bibinfo {volume} {127}},\ \bibinfo {pages} {151301} (\bibinfo {year} {2021})},\ \Eprint {https://arxiv.org/abs/2110.00483} {arXiv:2110.00483 [astro-ph.CO]} \BibitemShut {NoStop}%
\bibitem [{\citenamefont {Tristram}\ \emph {et~al.}(2022)\citenamefont {Tristram} \emph {et~al.}}]{Tristram:2021tvh}%
  \BibitemOpen
  \bibfield  {author} {\bibinfo {author} {\bibfnamefont {M.}~\bibnamefont {Tristram}} \emph {et~al.},\ }\bibfield  {title} {\bibinfo {title} {{Improved limits on the tensor-to-scalar ratio using BICEP and Planck data}},\ }\href {https://doi.org/10.1103/PhysRevD.105.083524} {\bibfield  {journal} {\bibinfo  {journal} {Phys. Rev. D}\ }\textbf {\bibinfo {volume} {105}},\ \bibinfo {pages} {083524} (\bibinfo {year} {2022})},\ \Eprint {https://arxiv.org/abs/2112.07961} {arXiv:2112.07961 [astro-ph.CO]} \BibitemShut {NoStop}%
\bibitem [{\citenamefont {Senatore}\ \emph {et~al.}(2014)\citenamefont {Senatore}, \citenamefont {Silverstein},\ and\ \citenamefont {Zaldarriaga}}]{Senatore:2011sp}%
  \BibitemOpen
  \bibfield  {author} {\bibinfo {author} {\bibfnamefont {L.}~\bibnamefont {Senatore}}, \bibinfo {author} {\bibfnamefont {E.}~\bibnamefont {Silverstein}},\ and\ \bibinfo {author} {\bibfnamefont {M.}~\bibnamefont {Zaldarriaga}},\ }\bibfield  {title} {\bibinfo {title} {{New Sources of Gravitational Waves during Inflation}},\ }\href {https://doi.org/10.1088/1475-7516/2014/08/016} {\bibfield  {journal} {\bibinfo  {journal} {JCAP}\ }\textbf {\bibinfo {volume} {08}},\ \bibinfo {pages} {016}},\ \Eprint {https://arxiv.org/abs/1109.0542} {arXiv:1109.0542 [hep-th]} \BibitemShut {NoStop}%
\bibitem [{\citenamefont {Barnaby}\ \emph {et~al.}(2012{\natexlab{a}})\citenamefont {Barnaby}, \citenamefont {Moxon}, \citenamefont {Namba}, \citenamefont {Peloso}, \citenamefont {Shiu},\ and\ \citenamefont {Zhou}}]{Barnaby:2012xt}%
  \BibitemOpen
  \bibfield  {author} {\bibinfo {author} {\bibfnamefont {N.}~\bibnamefont {Barnaby}}, \bibinfo {author} {\bibfnamefont {J.}~\bibnamefont {Moxon}}, \bibinfo {author} {\bibfnamefont {R.}~\bibnamefont {Namba}}, \bibinfo {author} {\bibfnamefont {M.}~\bibnamefont {Peloso}}, \bibinfo {author} {\bibfnamefont {G.}~\bibnamefont {Shiu}},\ and\ \bibinfo {author} {\bibfnamefont {P.}~\bibnamefont {Zhou}},\ }\bibfield  {title} {\bibinfo {title} {{Gravity waves and non-Gaussian features from particle production in a sector gravitationally coupled to the inflaton}},\ }\href {https://doi.org/10.1103/PhysRevD.86.103508} {\bibfield  {journal} {\bibinfo  {journal} {Phys. Rev. D}\ }\textbf {\bibinfo {volume} {86}},\ \bibinfo {pages} {103508} (\bibinfo {year} {2012}{\natexlab{a}})},\ \Eprint {https://arxiv.org/abs/1206.6117} {arXiv:1206.6117 [astro-ph.CO]} \BibitemShut {NoStop}%
\bibitem [{\citenamefont {Biagetti}\ \emph {et~al.}(2013)\citenamefont {Biagetti}, \citenamefont {Fasiello},\ and\ \citenamefont {Riotto}}]{Biagetti:2013kwa}%
  \BibitemOpen
  \bibfield  {author} {\bibinfo {author} {\bibfnamefont {M.}~\bibnamefont {Biagetti}}, \bibinfo {author} {\bibfnamefont {M.}~\bibnamefont {Fasiello}},\ and\ \bibinfo {author} {\bibfnamefont {A.}~\bibnamefont {Riotto}},\ }\bibfield  {title} {\bibinfo {title} {{Enhancing Inflationary Tensor Modes through Spectator Fields}},\ }\href {https://doi.org/10.1103/PhysRevD.88.103518} {\bibfield  {journal} {\bibinfo  {journal} {Phys. Rev. D}\ }\textbf {\bibinfo {volume} {88}},\ \bibinfo {pages} {103518} (\bibinfo {year} {2013})},\ \Eprint {https://arxiv.org/abs/1305.7241} {arXiv:1305.7241 [astro-ph.CO]} \BibitemShut {NoStop}%
\bibitem [{\citenamefont {Biagetti}\ \emph {et~al.}(2015)\citenamefont {Biagetti}, \citenamefont {Dimastrogiovanni}, \citenamefont {Fasiello},\ and\ \citenamefont {Peloso}}]{Biagetti:2014asa}%
  \BibitemOpen
  \bibfield  {author} {\bibinfo {author} {\bibfnamefont {M.}~\bibnamefont {Biagetti}}, \bibinfo {author} {\bibfnamefont {E.}~\bibnamefont {Dimastrogiovanni}}, \bibinfo {author} {\bibfnamefont {M.}~\bibnamefont {Fasiello}},\ and\ \bibinfo {author} {\bibfnamefont {M.}~\bibnamefont {Peloso}},\ }\bibfield  {title} {\bibinfo {title} {{Gravitational Waves and Scalar Perturbations from Spectator Fields}},\ }\href {https://doi.org/10.1088/1475-7516/2015/04/011} {\bibfield  {journal} {\bibinfo  {journal} {JCAP}\ }\textbf {\bibinfo {volume} {04}},\ \bibinfo {pages} {011}},\ \Eprint {https://arxiv.org/abs/1411.3029} {arXiv:1411.3029 [astro-ph.CO]} \BibitemShut {NoStop}%
\bibitem [{\citenamefont {Mirbabayi}\ \emph {et~al.}(2015)\citenamefont {Mirbabayi}, \citenamefont {Senatore}, \citenamefont {Silverstein},\ and\ \citenamefont {Zaldarriaga}}]{Mirbabayi:2014jqa}%
  \BibitemOpen
  \bibfield  {author} {\bibinfo {author} {\bibfnamefont {M.}~\bibnamefont {Mirbabayi}}, \bibinfo {author} {\bibfnamefont {L.}~\bibnamefont {Senatore}}, \bibinfo {author} {\bibfnamefont {E.}~\bibnamefont {Silverstein}},\ and\ \bibinfo {author} {\bibfnamefont {M.}~\bibnamefont {Zaldarriaga}},\ }\bibfield  {title} {\bibinfo {title} {{Gravitational Waves and the Scale of Inflation}},\ }\href {https://doi.org/10.1103/PhysRevD.91.063518} {\bibfield  {journal} {\bibinfo  {journal} {Phys. Rev. D}\ }\textbf {\bibinfo {volume} {91}},\ \bibinfo {pages} {063518} (\bibinfo {year} {2015})},\ \Eprint {https://arxiv.org/abs/1412.0665} {arXiv:1412.0665 [hep-th]} \BibitemShut {NoStop}%
\bibitem [{\citenamefont {Fujita}\ \emph {et~al.}(2015)\citenamefont {Fujita}, \citenamefont {Yokoyama},\ and\ \citenamefont {Yokoyama}}]{Fujita:2014oba}%
  \BibitemOpen
  \bibfield  {author} {\bibinfo {author} {\bibfnamefont {T.}~\bibnamefont {Fujita}}, \bibinfo {author} {\bibfnamefont {J.}~\bibnamefont {Yokoyama}},\ and\ \bibinfo {author} {\bibfnamefont {S.}~\bibnamefont {Yokoyama}},\ }\bibfield  {title} {\bibinfo {title} {{Can a spectator scalar field enhance inflationary tensor mode?}},\ }\href {https://doi.org/10.1093/ptep/ptv037} {\bibfield  {journal} {\bibinfo  {journal} {PTEP}\ }\textbf {\bibinfo {volume} {2015}},\ \bibinfo {pages} {043E01} (\bibinfo {year} {2015})},\ \Eprint {https://arxiv.org/abs/1411.3658} {arXiv:1411.3658 [astro-ph.CO]} \BibitemShut {NoStop}%
\bibitem [{\citenamefont {Yu}\ \emph {et~al.}(2023)\citenamefont {Yu}, \citenamefont {Fu},\ and\ \citenamefont {Guo}}]{Yu:2023ity}%
  \BibitemOpen
  \bibfield  {author} {\bibinfo {author} {\bibfnamefont {Z.}~\bibnamefont {Yu}}, \bibinfo {author} {\bibfnamefont {C.}~\bibnamefont {Fu}},\ and\ \bibinfo {author} {\bibfnamefont {Z.-K.}\ \bibnamefont {Guo}},\ }\bibfield  {title} {\bibinfo {title} {{Particle production during inflation with a nonminimally coupled spectator scalar field}},\ }\href {https://doi.org/10.1103/PhysRevD.108.123509} {\bibfield  {journal} {\bibinfo  {journal} {Phys. Rev. D}\ }\textbf {\bibinfo {volume} {108}},\ \bibinfo {pages} {123509} (\bibinfo {year} {2023})},\ \Eprint {https://arxiv.org/abs/2307.03120} {arXiv:2307.03120 [gr-qc]} \BibitemShut {NoStop}%
\bibitem [{\citenamefont {Ananda}\ \emph {et~al.}(2007)\citenamefont {Ananda}, \citenamefont {Clarkson},\ and\ \citenamefont {Wands}}]{Ananda:2006af}%
  \BibitemOpen
  \bibfield  {author} {\bibinfo {author} {\bibfnamefont {K.~N.}\ \bibnamefont {Ananda}}, \bibinfo {author} {\bibfnamefont {C.}~\bibnamefont {Clarkson}},\ and\ \bibinfo {author} {\bibfnamefont {D.}~\bibnamefont {Wands}},\ }\bibfield  {title} {\bibinfo {title} {{The Cosmological gravitational wave background from primordial density perturbations}},\ }\href {https://doi.org/10.1103/PhysRevD.75.123518} {\bibfield  {journal} {\bibinfo  {journal} {Phys. Rev. D}\ }\textbf {\bibinfo {volume} {75}},\ \bibinfo {pages} {123518} (\bibinfo {year} {2007})},\ \Eprint {https://arxiv.org/abs/gr-qc/0612013} {arXiv:gr-qc/0612013} \BibitemShut {NoStop}%
\bibitem [{\citenamefont {Baumann}\ \emph {et~al.}(2007)\citenamefont {Baumann}, \citenamefont {Steinhardt}, \citenamefont {Takahashi},\ and\ \citenamefont {Ichiki}}]{Baumann:2007zm}%
  \BibitemOpen
  \bibfield  {author} {\bibinfo {author} {\bibfnamefont {D.}~\bibnamefont {Baumann}}, \bibinfo {author} {\bibfnamefont {P.~J.}\ \bibnamefont {Steinhardt}}, \bibinfo {author} {\bibfnamefont {K.}~\bibnamefont {Takahashi}},\ and\ \bibinfo {author} {\bibfnamefont {K.}~\bibnamefont {Ichiki}},\ }\bibfield  {title} {\bibinfo {title} {{Gravitational Wave Spectrum Induced by Primordial Scalar Perturbations}},\ }\href {https://doi.org/10.1103/PhysRevD.76.084019} {\bibfield  {journal} {\bibinfo  {journal} {Phys. Rev. D}\ }\textbf {\bibinfo {volume} {76}},\ \bibinfo {pages} {084019} (\bibinfo {year} {2007})},\ \Eprint {https://arxiv.org/abs/hep-th/0703290} {arXiv:hep-th/0703290} \BibitemShut {NoStop}%
\bibitem [{\citenamefont {Kohri}\ and\ \citenamefont {Terada}(2018)}]{Kohri:2018awv}%
  \BibitemOpen
  \bibfield  {author} {\bibinfo {author} {\bibfnamefont {K.}~\bibnamefont {Kohri}}\ and\ \bibinfo {author} {\bibfnamefont {T.}~\bibnamefont {Terada}},\ }\bibfield  {title} {\bibinfo {title} {{Semianalytic calculation of gravitational wave spectrum nonlinearly induced from primordial curvature perturbations}},\ }\href {https://doi.org/10.1103/PhysRevD.97.123532} {\bibfield  {journal} {\bibinfo  {journal} {Phys. Rev. D}\ }\textbf {\bibinfo {volume} {97}},\ \bibinfo {pages} {123532} (\bibinfo {year} {2018})},\ \Eprint {https://arxiv.org/abs/1804.08577} {arXiv:1804.08577 [gr-qc]} \BibitemShut {NoStop}%
\bibitem [{\citenamefont {Fu}\ \emph {et~al.}(2020)\citenamefont {Fu}, \citenamefont {Wu},\ and\ \citenamefont {Yu}}]{Fu:2019vqc}%
  \BibitemOpen
  \bibfield  {author} {\bibinfo {author} {\bibfnamefont {C.}~\bibnamefont {Fu}}, \bibinfo {author} {\bibfnamefont {P.}~\bibnamefont {Wu}},\ and\ \bibinfo {author} {\bibfnamefont {H.}~\bibnamefont {Yu}},\ }\bibfield  {title} {\bibinfo {title} {{Scalar induced gravitational waves in inflation with gravitationally enhanced friction}},\ }\href {https://doi.org/10.1103/PhysRevD.101.023529} {\bibfield  {journal} {\bibinfo  {journal} {Phys. Rev. D}\ }\textbf {\bibinfo {volume} {101}},\ \bibinfo {pages} {023529} (\bibinfo {year} {2020})},\ \Eprint {https://arxiv.org/abs/1912.05927} {arXiv:1912.05927 [astro-ph.CO]} \BibitemShut {NoStop}%
\bibitem [{\citenamefont {Dom\`enech}(2020)}]{Domenech:2019quo}%
  \BibitemOpen
  \bibfield  {author} {\bibinfo {author} {\bibfnamefont {G.}~\bibnamefont {Dom\`enech}},\ }\bibfield  {title} {\bibinfo {title} {{Induced gravitational waves in a general cosmological background}},\ }\href {https://doi.org/10.1142/S0218271820500285} {\bibfield  {journal} {\bibinfo  {journal} {Int. J. Mod. Phys. D}\ }\textbf {\bibinfo {volume} {29}},\ \bibinfo {pages} {2050028} (\bibinfo {year} {2020})},\ \Eprint {https://arxiv.org/abs/1912.05583} {arXiv:1912.05583 [gr-qc]} \BibitemShut {NoStop}%
\bibitem [{\citenamefont {Dom\`enech}\ \emph {et~al.}(2020)\citenamefont {Dom\`enech}, \citenamefont {Pi},\ and\ \citenamefont {Sasaki}}]{Domenech:2020kqm}%
  \BibitemOpen
  \bibfield  {author} {\bibinfo {author} {\bibfnamefont {G.}~\bibnamefont {Dom\`enech}}, \bibinfo {author} {\bibfnamefont {S.}~\bibnamefont {Pi}},\ and\ \bibinfo {author} {\bibfnamefont {M.}~\bibnamefont {Sasaki}},\ }\bibfield  {title} {\bibinfo {title} {{Induced gravitational waves as a probe of thermal history of the universe}},\ }\href {https://doi.org/10.1088/1475-7516/2020/08/017} {\bibfield  {journal} {\bibinfo  {journal} {JCAP}\ }\textbf {\bibinfo {volume} {08}},\ \bibinfo {pages} {017}},\ \Eprint {https://arxiv.org/abs/2005.12314} {arXiv:2005.12314 [gr-qc]} \BibitemShut {NoStop}%
\bibitem [{\citenamefont {Pi}\ and\ \citenamefont {Sasaki}(2020)}]{Pi:2020otn}%
  \BibitemOpen
  \bibfield  {author} {\bibinfo {author} {\bibfnamefont {S.}~\bibnamefont {Pi}}\ and\ \bibinfo {author} {\bibfnamefont {M.}~\bibnamefont {Sasaki}},\ }\bibfield  {title} {\bibinfo {title} {{Gravitational Waves Induced by Scalar Perturbations with a Lognormal Peak}},\ }\href {https://doi.org/10.1088/1475-7516/2020/09/037} {\bibfield  {journal} {\bibinfo  {journal} {JCAP}\ }\textbf {\bibinfo {volume} {09}},\ \bibinfo {pages} {037}},\ \Eprint {https://arxiv.org/abs/2005.12306} {arXiv:2005.12306 [gr-qc]} \BibitemShut {NoStop}%
\bibitem [{\citenamefont {Cai}\ \emph {et~al.}(2019)\citenamefont {Cai}, \citenamefont {Pi},\ and\ \citenamefont {Sasaki}}]{Cai:2018dig}%
  \BibitemOpen
  \bibfield  {author} {\bibinfo {author} {\bibfnamefont {R.-g.}\ \bibnamefont {Cai}}, \bibinfo {author} {\bibfnamefont {S.}~\bibnamefont {Pi}},\ and\ \bibinfo {author} {\bibfnamefont {M.}~\bibnamefont {Sasaki}},\ }\bibfield  {title} {\bibinfo {title} {{Gravitational Waves Induced by non-Gaussian Scalar Perturbations}},\ }\href {https://doi.org/10.1103/PhysRevLett.122.201101} {\bibfield  {journal} {\bibinfo  {journal} {Phys. Rev. Lett.}\ }\textbf {\bibinfo {volume} {122}},\ \bibinfo {pages} {201101} (\bibinfo {year} {2019})},\ \Eprint {https://arxiv.org/abs/1810.11000} {arXiv:1810.11000 [astro-ph.CO]} \BibitemShut {NoStop}%
\bibitem [{\citenamefont {Antoniadis}\ \emph {et~al.}(2023)\citenamefont {Antoniadis} \emph {et~al.}}]{EPTA:2023fyk}%
  \BibitemOpen
  \bibfield  {author} {\bibinfo {author} {\bibfnamefont {J.}~\bibnamefont {Antoniadis}} \emph {et~al.} (\bibinfo {collaboration} {EPTA, InPTA:}),\ }\bibfield  {title} {\bibinfo {title} {{The second data release from the European Pulsar Timing Array - III. Search for gravitational wave signals}},\ }\href {https://doi.org/10.1051/0004-6361/202346844} {\bibfield  {journal} {\bibinfo  {journal} {Astron. Astrophys.}\ }\textbf {\bibinfo {volume} {678}},\ \bibinfo {pages} {A50} (\bibinfo {year} {2023})},\ \Eprint {https://arxiv.org/abs/2306.16214} {arXiv:2306.16214 [astro-ph.HE]} \BibitemShut {NoStop}%
\bibitem [{\citenamefont {Antoniadis}\ \emph {et~al.}(2024{\natexlab{a}})\citenamefont {Antoniadis} \emph {et~al.}}]{EPTA:2023gyr}%
  \BibitemOpen
  \bibfield  {author} {\bibinfo {author} {\bibfnamefont {J.}~\bibnamefont {Antoniadis}} \emph {et~al.} (\bibinfo {collaboration} {EPTA, InPTA}),\ }\bibfield  {title} {\bibinfo {title} {{The second data release from the European Pulsar Timing Array - V. Search for continuous gravitational wave signals}},\ }\href {https://doi.org/10.1051/0004-6361/202348568} {\bibfield  {journal} {\bibinfo  {journal} {Astron. Astrophys.}\ }\textbf {\bibinfo {volume} {690}},\ \bibinfo {pages} {A118} (\bibinfo {year} {2024}{\natexlab{a}})},\ \Eprint {https://arxiv.org/abs/2306.16226} {arXiv:2306.16226 [astro-ph.HE]} \BibitemShut {NoStop}%
\bibitem [{\citenamefont {Antoniadis}\ \emph {et~al.}(2024{\natexlab{b}})\citenamefont {Antoniadis} \emph {et~al.}}]{EPTA:2023xxk}%
  \BibitemOpen
  \bibfield  {author} {\bibinfo {author} {\bibfnamefont {J.}~\bibnamefont {Antoniadis}} \emph {et~al.} (\bibinfo {collaboration} {EPTA, InPTA}),\ }\bibfield  {title} {\bibinfo {title} {{The second data release from the European Pulsar Timing Array - IV. Implications for massive black holes, dark matter, and the early Universe}},\ }\href {https://doi.org/10.1051/0004-6361/202347433} {\bibfield  {journal} {\bibinfo  {journal} {Astron. Astrophys.}\ }\textbf {\bibinfo {volume} {685}},\ \bibinfo {pages} {A94} (\bibinfo {year} {2024}{\natexlab{b}})},\ \Eprint {https://arxiv.org/abs/2306.16227} {arXiv:2306.16227 [astro-ph.CO]} \BibitemShut {NoStop}%
\bibitem [{\citenamefont {Arzoumanian}\ \emph {et~al.}(2016)\citenamefont {Arzoumanian} \emph {et~al.}}]{NANOGrav:2015aud}%
  \BibitemOpen
  \bibfield  {author} {\bibinfo {author} {\bibfnamefont {Z.}~\bibnamefont {Arzoumanian}} \emph {et~al.} (\bibinfo {collaboration} {NANOGrav}),\ }\bibfield  {title} {\bibinfo {title} {{The NANOGrav Nine-year Data Set: Limits on the Isotropic Stochastic Gravitational Wave Background}},\ }\href {https://doi.org/10.3847/0004-637X/821/1/13} {\bibfield  {journal} {\bibinfo  {journal} {Astrophys. J.}\ }\textbf {\bibinfo {volume} {821}},\ \bibinfo {pages} {13} (\bibinfo {year} {2016})},\ \Eprint {https://arxiv.org/abs/1508.03024} {arXiv:1508.03024 [astro-ph.GA]} \BibitemShut {NoStop}%
\bibitem [{\citenamefont {Arzoumanian}\ \emph {et~al.}(2018)\citenamefont {Arzoumanian} \emph {et~al.}}]{NANOGRAV:2018hou}%
  \BibitemOpen
  \bibfield  {author} {\bibinfo {author} {\bibfnamefont {Z.}~\bibnamefont {Arzoumanian}} \emph {et~al.} (\bibinfo {collaboration} {NANOGRAV}),\ }\bibfield  {title} {\bibinfo {title} {{The NANOGrav 11-year Data Set: Pulsar-timing Constraints On The Stochastic Gravitational-wave Background}},\ }\href {https://doi.org/10.3847/1538-4357/aabd3b} {\bibfield  {journal} {\bibinfo  {journal} {Astrophys. J.}\ }\textbf {\bibinfo {volume} {859}},\ \bibinfo {pages} {47} (\bibinfo {year} {2018})},\ \Eprint {https://arxiv.org/abs/1801.02617} {arXiv:1801.02617 [astro-ph.HE]} \BibitemShut {NoStop}%
\bibitem [{\citenamefont {Carilli}\ and\ \citenamefont {Rawlings}(2004)}]{Carilli:2004nx}%
  \BibitemOpen
  \bibfield  {author} {\bibinfo {author} {\bibfnamefont {C.~L.}\ \bibnamefont {Carilli}}\ and\ \bibinfo {author} {\bibfnamefont {S.}~\bibnamefont {Rawlings}},\ }\bibfield  {title} {\bibinfo {title} {{Science with the Square Kilometer Array: Motivation, key science projects, standards and assumptions}},\ }\href {https://doi.org/10.1016/j.newar.2004.09.001} {\bibfield  {journal} {\bibinfo  {journal} {New Astron. Rev.}\ }\textbf {\bibinfo {volume} {48}},\ \bibinfo {pages} {979} (\bibinfo {year} {2004})},\ \Eprint {https://arxiv.org/abs/astro-ph/0409274} {arXiv:astro-ph/0409274} \BibitemShut {NoStop}%
\bibitem [{\citenamefont {Janssen}\ \emph {et~al.}(2015)\citenamefont {Janssen} \emph {et~al.}}]{Janssen:2014dka}%
  \BibitemOpen
  \bibfield  {author} {\bibinfo {author} {\bibfnamefont {G.}~\bibnamefont {Janssen}} \emph {et~al.},\ }\bibfield  {title} {\bibinfo {title} {{Gravitational wave astronomy with the SKA}},\ }\href {https://doi.org/10.22323/1.215.0037} {\bibfield  {journal} {\bibinfo  {journal} {PoS}\ }\textbf {\bibinfo {volume} {AASKA14}},\ \bibinfo {pages} {037} (\bibinfo {year} {2015})},\ \Eprint {https://arxiv.org/abs/1501.00127} {arXiv:1501.00127 [astro-ph.IM]} \BibitemShut {NoStop}%
\bibitem [{\citenamefont {Amaro-Seoane}\ \emph {et~al.}(2017)\citenamefont {Amaro-Seoane} \emph {et~al.}}]{LISA:2017pwj}%
  \BibitemOpen
  \bibfield  {author} {\bibinfo {author} {\bibfnamefont {P.}~\bibnamefont {Amaro-Seoane}} \emph {et~al.} (\bibinfo {collaboration} {LISA}),\ }\bibfield  {title} {\bibinfo {title} {{Laser Interferometer Space Antenna}},\ }\href@noop {} {\  (\bibinfo {year} {2017})},\ \Eprint {https://arxiv.org/abs/1702.00786} {arXiv:1702.00786 [astro-ph.IM]} \BibitemShut {NoStop}%
\bibitem [{\citenamefont {Ruan}\ \emph {et~al.}(2020)\citenamefont {Ruan}, \citenamefont {Guo}, \citenamefont {Cai},\ and\ \citenamefont {Zhang}}]{Ruan:2018tsw}%
  \BibitemOpen
  \bibfield  {author} {\bibinfo {author} {\bibfnamefont {W.-H.}\ \bibnamefont {Ruan}}, \bibinfo {author} {\bibfnamefont {Z.-K.}\ \bibnamefont {Guo}}, \bibinfo {author} {\bibfnamefont {R.-G.}\ \bibnamefont {Cai}},\ and\ \bibinfo {author} {\bibfnamefont {Y.-Z.}\ \bibnamefont {Zhang}},\ }\bibfield  {title} {\bibinfo {title} {{Taiji program: Gravitational-wave sources}},\ }\href {https://doi.org/10.1142/S0217751X2050075X} {\bibfield  {journal} {\bibinfo  {journal} {Int. J. Mod. Phys. A}\ }\textbf {\bibinfo {volume} {35}},\ \bibinfo {pages} {2050075} (\bibinfo {year} {2020})},\ \Eprint {https://arxiv.org/abs/1807.09495} {arXiv:1807.09495 [gr-qc]} \BibitemShut {NoStop}%
\bibitem [{\citenamefont {Harry}(2010)}]{Harry:2010zz}%
  \BibitemOpen
  \bibfield  {author} {\bibinfo {author} {\bibfnamefont {G.~M.}\ \bibnamefont {Harry}} (\bibinfo {collaboration} {LIGO Scientific}),\ }\bibfield  {title} {\bibinfo {title} {{Advanced LIGO: The next generation of gravitational wave detectors}},\ }\href {https://doi.org/10.1088/0264-9381/27/8/084006} {\bibfield  {journal} {\bibinfo  {journal} {Class. Quant. Grav.}\ }\textbf {\bibinfo {volume} {27}},\ \bibinfo {pages} {084006} (\bibinfo {year} {2010})}\BibitemShut {NoStop}%
\bibitem [{\citenamefont {Acernese}\ \emph {et~al.}(2015)\citenamefont {Acernese} \emph {et~al.}}]{VIRGO:2014yos}%
  \BibitemOpen
  \bibfield  {author} {\bibinfo {author} {\bibfnamefont {F.}~\bibnamefont {Acernese}} \emph {et~al.} (\bibinfo {collaboration} {VIRGO}),\ }\bibfield  {title} {\bibinfo {title} {{Advanced Virgo: a second-generation interferometric gravitational wave detector}},\ }\href {https://doi.org/10.1088/0264-9381/32/2/024001} {\bibfield  {journal} {\bibinfo  {journal} {Class. Quant. Grav.}\ }\textbf {\bibinfo {volume} {32}},\ \bibinfo {pages} {024001} (\bibinfo {year} {2015})},\ \Eprint {https://arxiv.org/abs/1408.3978} {arXiv:1408.3978 [gr-qc]} \BibitemShut {NoStop}%
\bibitem [{\citenamefont {Barnaby}\ and\ \citenamefont {Peloso}(2011)}]{Barnaby:2010vf}%
  \BibitemOpen
  \bibfield  {author} {\bibinfo {author} {\bibfnamefont {N.}~\bibnamefont {Barnaby}}\ and\ \bibinfo {author} {\bibfnamefont {M.}~\bibnamefont {Peloso}},\ }\bibfield  {title} {\bibinfo {title} {{Large Nongaussianity in Axion Inflation}},\ }\href {https://doi.org/10.1103/PhysRevLett.106.181301} {\bibfield  {journal} {\bibinfo  {journal} {Phys. Rev. Lett.}\ }\textbf {\bibinfo {volume} {106}},\ \bibinfo {pages} {181301} (\bibinfo {year} {2011})},\ \Eprint {https://arxiv.org/abs/1011.1500} {arXiv:1011.1500 [hep-ph]} \BibitemShut {NoStop}%
\bibitem [{\citenamefont {Sorbo}(2011)}]{Sorbo:2011rz}%
  \BibitemOpen
  \bibfield  {author} {\bibinfo {author} {\bibfnamefont {L.}~\bibnamefont {Sorbo}},\ }\bibfield  {title} {\bibinfo {title} {{Parity violation in the Cosmic Microwave Background from a pseudoscalar inflaton}},\ }\href {https://doi.org/10.1088/1475-7516/2011/06/003} {\bibfield  {journal} {\bibinfo  {journal} {JCAP}\ }\textbf {\bibinfo {volume} {06}},\ \bibinfo {pages} {003}},\ \Eprint {https://arxiv.org/abs/1101.1525} {arXiv:1101.1525 [astro-ph.CO]} \BibitemShut {NoStop}%
\bibitem [{\citenamefont {Meerburg}\ and\ \citenamefont {Pajer}(2013)}]{Meerburg:2012id}%
  \BibitemOpen
  \bibfield  {author} {\bibinfo {author} {\bibfnamefont {P.~D.}\ \bibnamefont {Meerburg}}\ and\ \bibinfo {author} {\bibfnamefont {E.}~\bibnamefont {Pajer}},\ }\bibfield  {title} {\bibinfo {title} {{Observational Constraints on Gauge Field Production in Axion Inflation}},\ }\href {https://doi.org/10.1088/1475-7516/2013/02/017} {\bibfield  {journal} {\bibinfo  {journal} {JCAP}\ }\textbf {\bibinfo {volume} {02}},\ \bibinfo {pages} {017}},\ \Eprint {https://arxiv.org/abs/1203.6076} {arXiv:1203.6076 [astro-ph.CO]} \BibitemShut {NoStop}%
\bibitem [{\citenamefont {Urban}(2013)}]{Urban:2013spa}%
  \BibitemOpen
  \bibfield  {author} {\bibinfo {author} {\bibfnamefont {F.~R.}\ \bibnamefont {Urban}},\ }\bibfield  {title} {\bibinfo {title} {{Pseudoscalar N-flation and axial coupling revisited}},\ }\href {https://doi.org/10.1103/PhysRevD.88.063525} {\bibfield  {journal} {\bibinfo  {journal} {Phys. Rev. D}\ }\textbf {\bibinfo {volume} {88}},\ \bibinfo {pages} {063525} (\bibinfo {year} {2013})},\ \Eprint {https://arxiv.org/abs/1307.5215} {arXiv:1307.5215 [astro-ph.CO]} \BibitemShut {NoStop}%
\bibitem [{\citenamefont {Barnaby}\ \emph {et~al.}(2012{\natexlab{b}})\citenamefont {Barnaby}, \citenamefont {Pajer},\ and\ \citenamefont {Peloso}}]{Barnaby:2011qe}%
  \BibitemOpen
  \bibfield  {author} {\bibinfo {author} {\bibfnamefont {N.}~\bibnamefont {Barnaby}}, \bibinfo {author} {\bibfnamefont {E.}~\bibnamefont {Pajer}},\ and\ \bibinfo {author} {\bibfnamefont {M.}~\bibnamefont {Peloso}},\ }\bibfield  {title} {\bibinfo {title} {{Gauge Field Production in Axion Inflation: Consequences for Monodromy, non-Gaussianity in the CMB, and Gravitational Waves at Interferometers}},\ }\href {https://doi.org/10.1103/PhysRevD.85.023525} {\bibfield  {journal} {\bibinfo  {journal} {Phys. Rev. D}\ }\textbf {\bibinfo {volume} {85}},\ \bibinfo {pages} {023525} (\bibinfo {year} {2012}{\natexlab{b}})},\ \Eprint {https://arxiv.org/abs/1110.3327} {arXiv:1110.3327 [astro-ph.CO]} \BibitemShut {NoStop}%
\bibitem [{\citenamefont {Zhou}\ \emph {et~al.}(2020)\citenamefont {Zhou}, \citenamefont {Jiang}, \citenamefont {Cai}, \citenamefont {Sasaki},\ and\ \citenamefont {Pi}}]{Zhou:2020kkf}%
  \BibitemOpen
  \bibfield  {author} {\bibinfo {author} {\bibfnamefont {Z.}~\bibnamefont {Zhou}}, \bibinfo {author} {\bibfnamefont {J.}~\bibnamefont {Jiang}}, \bibinfo {author} {\bibfnamefont {Y.-F.}\ \bibnamefont {Cai}}, \bibinfo {author} {\bibfnamefont {M.}~\bibnamefont {Sasaki}},\ and\ \bibinfo {author} {\bibfnamefont {S.}~\bibnamefont {Pi}},\ }\bibfield  {title} {\bibinfo {title} {{Primordial black holes and gravitational waves from resonant amplification during inflation}},\ }\href {https://doi.org/10.1103/PhysRevD.102.103527} {\bibfield  {journal} {\bibinfo  {journal} {Phys. Rev. D}\ }\textbf {\bibinfo {volume} {102}},\ \bibinfo {pages} {103527} (\bibinfo {year} {2020})},\ \Eprint {https://arxiv.org/abs/2010.03537} {arXiv:2010.03537 [astro-ph.CO]} \BibitemShut {NoStop}%
\bibitem [{\citenamefont {Peng}\ \emph {et~al.}(2021)\citenamefont {Peng}, \citenamefont {Fu}, \citenamefont {Liu}, \citenamefont {Guo},\ and\ \citenamefont {Cai}}]{Peng:2021zon}%
  \BibitemOpen
  \bibfield  {author} {\bibinfo {author} {\bibfnamefont {Z.-Z.}\ \bibnamefont {Peng}}, \bibinfo {author} {\bibfnamefont {C.}~\bibnamefont {Fu}}, \bibinfo {author} {\bibfnamefont {J.}~\bibnamefont {Liu}}, \bibinfo {author} {\bibfnamefont {Z.-K.}\ \bibnamefont {Guo}},\ and\ \bibinfo {author} {\bibfnamefont {R.-G.}\ \bibnamefont {Cai}},\ }\bibfield  {title} {\bibinfo {title} {{Gravitational waves from resonant amplification of curvature perturbations during inflation}},\ }\href {https://doi.org/10.1088/1475-7516/2021/10/050} {\bibfield  {journal} {\bibinfo  {journal} {JCAP}\ }\textbf {\bibinfo {volume} {10}},\ \bibinfo {pages} {050}},\ \Eprint {https://arxiv.org/abs/2106.11816} {arXiv:2106.11816 [astro-ph.CO]} \BibitemShut {NoStop}%
\bibitem [{\citenamefont {Barnaby}\ \emph {et~al.}(2011)\citenamefont {Barnaby}, \citenamefont {Namba},\ and\ \citenamefont {Peloso}}]{Barnaby:2011vw}%
  \BibitemOpen
  \bibfield  {author} {\bibinfo {author} {\bibfnamefont {N.}~\bibnamefont {Barnaby}}, \bibinfo {author} {\bibfnamefont {R.}~\bibnamefont {Namba}},\ and\ \bibinfo {author} {\bibfnamefont {M.}~\bibnamefont {Peloso}},\ }\bibfield  {title} {\bibinfo {title} {{Phenomenology of a Pseudo-Scalar Inflaton: Naturally Large Nongaussianity}},\ }\href {https://doi.org/10.1088/1475-7516/2011/04/009} {\bibfield  {journal} {\bibinfo  {journal} {JCAP}\ }\textbf {\bibinfo {volume} {04}},\ \bibinfo {pages} {009}},\ \Eprint {https://arxiv.org/abs/1102.4333} {arXiv:1102.4333 [astro-ph.CO]} \BibitemShut {NoStop}%
\bibitem [{\citenamefont {Cook}\ and\ \citenamefont {Sorbo}(2012)}]{Cook:2011hg}%
  \BibitemOpen
  \bibfield  {author} {\bibinfo {author} {\bibfnamefont {J.~L.}\ \bibnamefont {Cook}}\ and\ \bibinfo {author} {\bibfnamefont {L.}~\bibnamefont {Sorbo}},\ }\bibfield  {title} {\bibinfo {title} {{Particle production during inflation and gravitational waves detectable by ground-based interferometers}},\ }\href {https://doi.org/10.1103/PhysRevD.85.023534} {\bibfield  {journal} {\bibinfo  {journal} {Phys. Rev. D}\ }\textbf {\bibinfo {volume} {85}},\ \bibinfo {pages} {023534} (\bibinfo {year} {2012})},\ \bibinfo {note} {[Erratum: Phys.Rev.D 86, 069901 (2012)]},\ \Eprint {https://arxiv.org/abs/1109.0022} {arXiv:1109.0022 [astro-ph.CO]} \BibitemShut {NoStop}%
\bibitem [{\citenamefont {Dimopoulos}\ and\ \citenamefont {Karciauskas}(2012)}]{Dimopoulos:2012av}%
  \BibitemOpen
  \bibfield  {author} {\bibinfo {author} {\bibfnamefont {K.}~\bibnamefont {Dimopoulos}}\ and\ \bibinfo {author} {\bibfnamefont {M.}~\bibnamefont {Karciauskas}},\ }\bibfield  {title} {\bibinfo {title} {{Parity Violating Statistical Anisotropy}},\ }\href {https://doi.org/10.1007/JHEP06(2012)040} {\bibfield  {journal} {\bibinfo  {journal} {JHEP}\ }\textbf {\bibinfo {volume} {06}},\ \bibinfo {pages} {040}},\ \Eprint {https://arxiv.org/abs/1203.0230} {arXiv:1203.0230 [hep-ph]} \BibitemShut {NoStop}%
\bibitem [{\citenamefont {Linde}\ \emph {et~al.}(2013)\citenamefont {Linde}, \citenamefont {Mooij},\ and\ \citenamefont {Pajer}}]{Linde:2012bt}%
  \BibitemOpen
  \bibfield  {author} {\bibinfo {author} {\bibfnamefont {A.}~\bibnamefont {Linde}}, \bibinfo {author} {\bibfnamefont {S.}~\bibnamefont {Mooij}},\ and\ \bibinfo {author} {\bibfnamefont {E.}~\bibnamefont {Pajer}},\ }\bibfield  {title} {\bibinfo {title} {{Gauge field production in supergravity inflation: Local non-Gaussianity and primordial black holes}},\ }\href {https://doi.org/10.1103/PhysRevD.87.103506} {\bibfield  {journal} {\bibinfo  {journal} {Phys. Rev. D}\ }\textbf {\bibinfo {volume} {87}},\ \bibinfo {pages} {103506} (\bibinfo {year} {2013})},\ \Eprint {https://arxiv.org/abs/1212.1693} {arXiv:1212.1693 [hep-th]} \BibitemShut {NoStop}%
\bibitem [{\citenamefont {Mukohyama}\ \emph {et~al.}(2014)\citenamefont {Mukohyama}, \citenamefont {Namba}, \citenamefont {Peloso},\ and\ \citenamefont {Shiu}}]{Mukohyama:2014gba}%
  \BibitemOpen
  \bibfield  {author} {\bibinfo {author} {\bibfnamefont {S.}~\bibnamefont {Mukohyama}}, \bibinfo {author} {\bibfnamefont {R.}~\bibnamefont {Namba}}, \bibinfo {author} {\bibfnamefont {M.}~\bibnamefont {Peloso}},\ and\ \bibinfo {author} {\bibfnamefont {G.}~\bibnamefont {Shiu}},\ }\bibfield  {title} {\bibinfo {title} {{Blue Tensor Spectrum from Particle Production during Inflation}},\ }\href {https://doi.org/10.1088/1475-7516/2014/08/036} {\bibfield  {journal} {\bibinfo  {journal} {JCAP}\ }\textbf {\bibinfo {volume} {08}},\ \bibinfo {pages} {036}},\ \Eprint {https://arxiv.org/abs/1405.0346} {arXiv:1405.0346 [astro-ph.CO]} \BibitemShut {NoStop}%
\bibitem [{\citenamefont {Namba}\ \emph {et~al.}(2016)\citenamefont {Namba}, \citenamefont {Peloso}, \citenamefont {Shiraishi}, \citenamefont {Sorbo},\ and\ \citenamefont {Unal}}]{Namba:2015gja}%
  \BibitemOpen
  \bibfield  {author} {\bibinfo {author} {\bibfnamefont {R.}~\bibnamefont {Namba}}, \bibinfo {author} {\bibfnamefont {M.}~\bibnamefont {Peloso}}, \bibinfo {author} {\bibfnamefont {M.}~\bibnamefont {Shiraishi}}, \bibinfo {author} {\bibfnamefont {L.}~\bibnamefont {Sorbo}},\ and\ \bibinfo {author} {\bibfnamefont {C.}~\bibnamefont {Unal}},\ }\bibfield  {title} {\bibinfo {title} {{Scale-dependent gravitational waves from a rolling axion}},\ }\href {https://doi.org/10.1088/1475-7516/2016/01/041} {\bibfield  {journal} {\bibinfo  {journal} {JCAP}\ }\textbf {\bibinfo {volume} {01}},\ \bibinfo {pages} {041}},\ \Eprint {https://arxiv.org/abs/1509.07521} {arXiv:1509.07521 [astro-ph.CO]} \BibitemShut {NoStop}%
\bibitem [{\citenamefont {Garcia-Bellido}\ \emph {et~al.}(2016)\citenamefont {Garcia-Bellido}, \citenamefont {Peloso},\ and\ \citenamefont {Unal}}]{Garcia-Bellido:2016dkw}%
  \BibitemOpen
  \bibfield  {author} {\bibinfo {author} {\bibfnamefont {J.}~\bibnamefont {Garcia-Bellido}}, \bibinfo {author} {\bibfnamefont {M.}~\bibnamefont {Peloso}},\ and\ \bibinfo {author} {\bibfnamefont {C.}~\bibnamefont {Unal}},\ }\bibfield  {title} {\bibinfo {title} {{Gravitational waves at interferometer scales and primordial black holes in axion inflation}},\ }\href {https://doi.org/10.1088/1475-7516/2016/12/031} {\bibfield  {journal} {\bibinfo  {journal} {JCAP}\ }\textbf {\bibinfo {volume} {12}},\ \bibinfo {pages} {031}},\ \Eprint {https://arxiv.org/abs/1610.03763} {arXiv:1610.03763 [astro-ph.CO]} \BibitemShut {NoStop}%
\bibitem [{\citenamefont {Agrawal}\ \emph {et~al.}(2018{\natexlab{a}})\citenamefont {Agrawal}, \citenamefont {Fujita},\ and\ \citenamefont {Komatsu}}]{Agrawal:2017awz}%
  \BibitemOpen
  \bibfield  {author} {\bibinfo {author} {\bibfnamefont {A.}~\bibnamefont {Agrawal}}, \bibinfo {author} {\bibfnamefont {T.}~\bibnamefont {Fujita}},\ and\ \bibinfo {author} {\bibfnamefont {E.}~\bibnamefont {Komatsu}},\ }\bibfield  {title} {\bibinfo {title} {{Large tensor non-Gaussianity from axion-gauge field dynamics}},\ }\href {https://doi.org/10.1103/PhysRevD.97.103526} {\bibfield  {journal} {\bibinfo  {journal} {Phys. Rev. D}\ }\textbf {\bibinfo {volume} {97}},\ \bibinfo {pages} {103526} (\bibinfo {year} {2018}{\natexlab{a}})},\ \Eprint {https://arxiv.org/abs/1707.03023} {arXiv:1707.03023 [astro-ph.CO]} \BibitemShut {NoStop}%
\bibitem [{\citenamefont {\"Ozsoy}(2018)}]{Ozsoy:2017blg}%
  \BibitemOpen
  \bibfield  {author} {\bibinfo {author} {\bibfnamefont {O.}~\bibnamefont {\"Ozsoy}},\ }\bibfield  {title} {\bibinfo {title} {{On Synthetic Gravitational Waves from Multi-field Inflation}},\ }\href {https://doi.org/10.1088/1475-7516/2018/04/062} {\bibfield  {journal} {\bibinfo  {journal} {JCAP}\ }\textbf {\bibinfo {volume} {04}},\ \bibinfo {pages} {062}},\ \Eprint {https://arxiv.org/abs/1712.01991} {arXiv:1712.01991 [astro-ph.CO]} \BibitemShut {NoStop}%
\bibitem [{\citenamefont {Agrawal}\ \emph {et~al.}(2018{\natexlab{b}})\citenamefont {Agrawal}, \citenamefont {Fujita},\ and\ \citenamefont {Komatsu}}]{Agrawal:2018mrg}%
  \BibitemOpen
  \bibfield  {author} {\bibinfo {author} {\bibfnamefont {A.}~\bibnamefont {Agrawal}}, \bibinfo {author} {\bibfnamefont {T.}~\bibnamefont {Fujita}},\ and\ \bibinfo {author} {\bibfnamefont {E.}~\bibnamefont {Komatsu}},\ }\bibfield  {title} {\bibinfo {title} {{Tensor Non-Gaussianity from Axion-Gauge-Fields Dynamics : Parameter Search}},\ }\href {https://doi.org/10.1088/1475-7516/2018/06/027} {\bibfield  {journal} {\bibinfo  {journal} {JCAP}\ }\textbf {\bibinfo {volume} {06}},\ \bibinfo {pages} {027}},\ \Eprint {https://arxiv.org/abs/1802.09284} {arXiv:1802.09284 [astro-ph.CO]} \BibitemShut {NoStop}%
\bibitem [{\citenamefont {Papageorgiou}\ \emph {et~al.}(2019)\citenamefont {Papageorgiou}, \citenamefont {Peloso},\ and\ \citenamefont {Unal}}]{Papageorgiou:2019ecb}%
  \BibitemOpen
  \bibfield  {author} {\bibinfo {author} {\bibfnamefont {A.}~\bibnamefont {Papageorgiou}}, \bibinfo {author} {\bibfnamefont {M.}~\bibnamefont {Peloso}},\ and\ \bibinfo {author} {\bibfnamefont {C.}~\bibnamefont {Unal}},\ }\bibfield  {title} {\bibinfo {title} {{Nonlinear perturbations from axion-gauge fields dynamics during inflation}},\ }\href {https://doi.org/10.1088/1475-7516/2019/07/004} {\bibfield  {journal} {\bibinfo  {journal} {JCAP}\ }\textbf {\bibinfo {volume} {07}},\ \bibinfo {pages} {004}},\ \Eprint {https://arxiv.org/abs/1904.01488} {arXiv:1904.01488 [astro-ph.CO]} \BibitemShut {NoStop}%
\bibitem [{\citenamefont {Campeti}\ \emph {et~al.}(2022)\citenamefont {Campeti}, \citenamefont {\"Ozsoy}, \citenamefont {Obata},\ and\ \citenamefont {Shiraishi}}]{Campeti:2022acx}%
  \BibitemOpen
  \bibfield  {author} {\bibinfo {author} {\bibfnamefont {P.}~\bibnamefont {Campeti}}, \bibinfo {author} {\bibfnamefont {O.}~\bibnamefont {\"Ozsoy}}, \bibinfo {author} {\bibfnamefont {I.}~\bibnamefont {Obata}},\ and\ \bibinfo {author} {\bibfnamefont {M.}~\bibnamefont {Shiraishi}},\ }\bibfield  {title} {\bibinfo {title} {{New constraints on axion-gauge field dynamics during inflation from Planck and BICEP/Keck data sets}},\ }\href {https://doi.org/10.1088/1475-7516/2022/07/039} {\bibfield  {journal} {\bibinfo  {journal} {JCAP}\ }\textbf {\bibinfo {volume} {07}}\bibfield  {number} {\bibinfo  {number} { (07)},\ \bibinfo {pages} {039}},\ }\Eprint {https://arxiv.org/abs/2203.03401} {arXiv:2203.03401 [astro-ph.CO]} \BibitemShut {NoStop}%
\bibitem [{\citenamefont {\"Ozsoy}\ \emph {et~al.}(2024)\citenamefont {\"Ozsoy}, \citenamefont {Papageorgiou},\ and\ \citenamefont {Fasiello}}]{Ozsoy:2024apn}%
  \BibitemOpen
  \bibfield  {author} {\bibinfo {author} {\bibfnamefont {O.}~\bibnamefont {\"Ozsoy}}, \bibinfo {author} {\bibfnamefont {A.}~\bibnamefont {Papageorgiou}},\ and\ \bibinfo {author} {\bibfnamefont {M.}~\bibnamefont {Fasiello}},\ }\bibfield  {title} {\bibinfo {title} {{Scale-dependent chirality as a smoking gun for Abelian gauge fields during inflation}},\ }\href@noop {} {\  (\bibinfo {year} {2024})},\ \Eprint {https://arxiv.org/abs/2405.14963} {arXiv:2405.14963 [astro-ph.CO]} \BibitemShut {NoStop}%
\bibitem [{\citenamefont {He}\ \emph {et~al.}(2025{\natexlab{a}})\citenamefont {He}, \citenamefont {Fu}, \citenamefont {Zhang},\ and\ \citenamefont {Guo}}]{He:2024bno}%
  \BibitemOpen
  \bibfield  {author} {\bibinfo {author} {\bibfnamefont {J.-F.}\ \bibnamefont {He}}, \bibinfo {author} {\bibfnamefont {C.}~\bibnamefont {Fu}}, \bibinfo {author} {\bibfnamefont {K.-G.}\ \bibnamefont {Zhang}},\ and\ \bibinfo {author} {\bibfnamefont {Z.-K.}\ \bibnamefont {Guo}},\ }\bibfield  {title} {\bibinfo {title} {{Gravitational waves from a gauge field nonminimally coupled to gravity}},\ }\href {https://doi.org/10.1103/PhysRevD.111.023536} {\bibfield  {journal} {\bibinfo  {journal} {Phys. Rev. D}\ }\textbf {\bibinfo {volume} {111}},\ \bibinfo {pages} {023536} (\bibinfo {year} {2025}{\natexlab{a}})},\ \Eprint {https://arxiv.org/abs/2409.08312} {arXiv:2409.08312 [astro-ph.CO]} \BibitemShut {NoStop}%
\bibitem [{\citenamefont {Alaei}\ \emph {et~al.}(2025)\citenamefont {Alaei}, \citenamefont {Bhattacharya},\ and\ \citenamefont {Kamali}}]{Alaei:2025ryv}%
  \BibitemOpen
  \bibfield  {author} {\bibinfo {author} {\bibfnamefont {A.}~\bibnamefont {Alaei}}, \bibinfo {author} {\bibfnamefont {S.}~\bibnamefont {Bhattacharya}},\ and\ \bibinfo {author} {\bibfnamefont {V.}~\bibnamefont {Kamali}},\ }\bibfield  {title} {\bibinfo {title} {{CMB constraints on $U(1)$ axion warm inflation}},\ }\href@noop {} {\  (\bibinfo {year} {2025})},\ \Eprint {https://arxiv.org/abs/2507.17438} {arXiv:2507.17438 [astro-ph.CO]} \BibitemShut {NoStop}%
\bibitem [{\citenamefont {Ferreira}\ and\ \citenamefont {Sloth}(2014)}]{Ferreira:2014zia}%
  \BibitemOpen
  \bibfield  {author} {\bibinfo {author} {\bibfnamefont {R.~Z.}\ \bibnamefont {Ferreira}}\ and\ \bibinfo {author} {\bibfnamefont {M.~S.}\ \bibnamefont {Sloth}},\ }\bibfield  {title} {\bibinfo {title} {{Universal Constraints on Axions from Inflation}},\ }\href {https://doi.org/10.1007/JHEP12(2014)139} {\bibfield  {journal} {\bibinfo  {journal} {JHEP}\ }\textbf {\bibinfo {volume} {12}},\ \bibinfo {pages} {139}},\ \Eprint {https://arxiv.org/abs/1409.5799} {arXiv:1409.5799 [hep-ph]} \BibitemShut {NoStop}%
\bibitem [{\citenamefont {Giar\`e}\ and\ \citenamefont {Melchiorri}(2021)}]{Giare:2020vhn}%
  \BibitemOpen
  \bibfield  {author} {\bibinfo {author} {\bibfnamefont {W.}~\bibnamefont {Giar\`e}}\ and\ \bibinfo {author} {\bibfnamefont {A.}~\bibnamefont {Melchiorri}},\ }\bibfield  {title} {\bibinfo {title} {{Probing the inflationary background of gravitational waves from large to small scales}},\ }\href {https://doi.org/10.1016/j.physletb.2021.136137} {\bibfield  {journal} {\bibinfo  {journal} {Phys. Lett. B}\ }\textbf {\bibinfo {volume} {815}},\ \bibinfo {pages} {136137} (\bibinfo {year} {2021})},\ \Eprint {https://arxiv.org/abs/2003.04783} {arXiv:2003.04783 [astro-ph.CO]} \BibitemShut {NoStop}%
\bibitem [{\citenamefont {Campeti}\ \emph {et~al.}(2021)\citenamefont {Campeti}, \citenamefont {Komatsu}, \citenamefont {Poletti},\ and\ \citenamefont {Baccigalupi}}]{Campeti:2020xwn}%
  \BibitemOpen
  \bibfield  {author} {\bibinfo {author} {\bibfnamefont {P.}~\bibnamefont {Campeti}}, \bibinfo {author} {\bibfnamefont {E.}~\bibnamefont {Komatsu}}, \bibinfo {author} {\bibfnamefont {D.}~\bibnamefont {Poletti}},\ and\ \bibinfo {author} {\bibfnamefont {C.}~\bibnamefont {Baccigalupi}},\ }\bibfield  {title} {\bibinfo {title} {{Measuring the spectrum of primordial gravitational waves with CMB, PTA and Laser Interferometers}},\ }\href {https://doi.org/10.1088/1475-7516/2021/01/012} {\bibfield  {journal} {\bibinfo  {journal} {JCAP}\ }\textbf {\bibinfo {volume} {01}},\ \bibinfo {pages} {012}},\ \Eprint {https://arxiv.org/abs/2007.04241} {arXiv:2007.04241 [astro-ph.CO]} \BibitemShut {NoStop}%
\bibitem [{\citenamefont {Auclair}\ \emph {et~al.}(2023)\citenamefont {Auclair} \emph {et~al.}}]{LISACosmologyWorkingGroup:2022jok}%
  \BibitemOpen
  \bibfield  {author} {\bibinfo {author} {\bibfnamefont {P.}~\bibnamefont {Auclair}} \emph {et~al.} (\bibinfo {collaboration} {LISA Cosmology Working Group}),\ }\bibfield  {title} {\bibinfo {title} {{Cosmology with the Laser Interferometer Space Antenna}},\ }\href {https://doi.org/10.1007/s41114-023-00045-2} {\bibfield  {journal} {\bibinfo  {journal} {Living Rev. Rel.}\ }\textbf {\bibinfo {volume} {26}},\ \bibinfo {pages} {5} (\bibinfo {year} {2023})},\ \Eprint {https://arxiv.org/abs/2204.05434} {arXiv:2204.05434 [astro-ph.CO]} \BibitemShut {NoStop}%
\bibitem [{\citenamefont {Garcia-Bellido}\ \emph {et~al.}(2024)\citenamefont {Garcia-Bellido}, \citenamefont {Papageorgiou}, \citenamefont {Peloso},\ and\ \citenamefont {Sorbo}}]{Garcia-Bellido:2023ser}%
  \BibitemOpen
  \bibfield  {author} {\bibinfo {author} {\bibfnamefont {J.}~\bibnamefont {Garcia-Bellido}}, \bibinfo {author} {\bibfnamefont {A.}~\bibnamefont {Papageorgiou}}, \bibinfo {author} {\bibfnamefont {M.}~\bibnamefont {Peloso}},\ and\ \bibinfo {author} {\bibfnamefont {L.}~\bibnamefont {Sorbo}},\ }\bibfield  {title} {\bibinfo {title} {{A flashing beacon in axion inflation: recurring bursts of gravitational waves in the strong backreaction regime}},\ }\href {https://doi.org/10.1088/1475-7516/2024/01/034} {\bibfield  {journal} {\bibinfo  {journal} {JCAP}\ }\textbf {\bibinfo {volume} {01}},\ \bibinfo {pages} {034}},\ \Eprint {https://arxiv.org/abs/2303.13425} {arXiv:2303.13425 [astro-ph.CO]} \BibitemShut {NoStop}%
\bibitem [{\citenamefont {Afzal}\ \emph {et~al.}(2023)\citenamefont {Afzal} \emph {et~al.}}]{NANOGrav:2023hvm}%
  \BibitemOpen
  \bibfield  {author} {\bibinfo {author} {\bibfnamefont {A.}~\bibnamefont {Afzal}} \emph {et~al.} (\bibinfo {collaboration} {NANOGrav}),\ }\bibfield  {title} {\bibinfo {title} {{The NANOGrav 15 yr Data Set: Search for Signals from New Physics}},\ }\href {https://doi.org/10.3847/2041-8213/acdc91} {\bibfield  {journal} {\bibinfo  {journal} {Astrophys. J. Lett.}\ }\textbf {\bibinfo {volume} {951}},\ \bibinfo {pages} {L11} (\bibinfo {year} {2023})},\ \Eprint {https://arxiv.org/abs/2306.16219} {arXiv:2306.16219 [astro-ph.HE]} \BibitemShut {NoStop}%
\bibitem [{\citenamefont {Figueroa}\ \emph {et~al.}(2024{\natexlab{a}})\citenamefont {Figueroa}, \citenamefont {Pieroni}, \citenamefont {Ricciardone},\ and\ \citenamefont {Simakachorn}}]{Figueroa:2023zhu}%
  \BibitemOpen
  \bibfield  {author} {\bibinfo {author} {\bibfnamefont {D.~G.}\ \bibnamefont {Figueroa}}, \bibinfo {author} {\bibfnamefont {M.}~\bibnamefont {Pieroni}}, \bibinfo {author} {\bibfnamefont {A.}~\bibnamefont {Ricciardone}},\ and\ \bibinfo {author} {\bibfnamefont {P.}~\bibnamefont {Simakachorn}},\ }\bibfield  {title} {\bibinfo {title} {{Cosmological Background Interpretation of Pulsar Timing Array Data}},\ }\href {https://doi.org/10.1103/PhysRevLett.132.171002} {\bibfield  {journal} {\bibinfo  {journal} {Phys. Rev. Lett.}\ }\textbf {\bibinfo {volume} {132}},\ \bibinfo {pages} {171002} (\bibinfo {year} {2024}{\natexlab{a}})},\ \Eprint {https://arxiv.org/abs/2307.02399} {arXiv:2307.02399 [astro-ph.CO]} \BibitemShut {NoStop}%
\bibitem [{\citenamefont {Unal}\ \emph {et~al.}(2024)\citenamefont {Unal}, \citenamefont {Papageorgiou},\ and\ \citenamefont {Obata}}]{Unal:2023srk}%
  \BibitemOpen
  \bibfield  {author} {\bibinfo {author} {\bibfnamefont {C.}~\bibnamefont {Unal}}, \bibinfo {author} {\bibfnamefont {A.}~\bibnamefont {Papageorgiou}},\ and\ \bibinfo {author} {\bibfnamefont {I.}~\bibnamefont {Obata}},\ }\bibfield  {title} {\bibinfo {title} {{Axion-gauge dynamics during inflation as the origin of pulsar timing array signals and primordial black holes}},\ }\href {https://doi.org/10.1016/j.physletb.2024.138873} {\bibfield  {journal} {\bibinfo  {journal} {Phys. Lett. B}\ }\textbf {\bibinfo {volume} {856}},\ \bibinfo {pages} {138873} (\bibinfo {year} {2024})},\ \Eprint {https://arxiv.org/abs/2307.02322} {arXiv:2307.02322 [astro-ph.CO]} \BibitemShut {NoStop}%
\bibitem [{\citenamefont {Niu}\ and\ \citenamefont {Rahat}(2023)}]{Niu:2023bsr}%
  \BibitemOpen
  \bibfield  {author} {\bibinfo {author} {\bibfnamefont {X.}~\bibnamefont {Niu}}\ and\ \bibinfo {author} {\bibfnamefont {M.~H.}\ \bibnamefont {Rahat}},\ }\bibfield  {title} {\bibinfo {title} {{NANOGrav signal from axion inflation}},\ }\href {https://doi.org/10.1103/PhysRevD.108.115023} {\bibfield  {journal} {\bibinfo  {journal} {Phys. Rev. D}\ }\textbf {\bibinfo {volume} {108}},\ \bibinfo {pages} {115023} (\bibinfo {year} {2023})},\ \Eprint {https://arxiv.org/abs/2307.01192} {arXiv:2307.01192 [hep-ph]} \BibitemShut {NoStop}%
\bibitem [{\citenamefont {Dimastrogiovanni}\ \emph {et~al.}(2023)\citenamefont {Dimastrogiovanni}, \citenamefont {Fasiello}, \citenamefont {Leedom}, \citenamefont {Putti},\ and\ \citenamefont {Westphal}}]{Dimastrogiovanni:2023juq}%
  \BibitemOpen
  \bibfield  {author} {\bibinfo {author} {\bibfnamefont {E.}~\bibnamefont {Dimastrogiovanni}}, \bibinfo {author} {\bibfnamefont {M.}~\bibnamefont {Fasiello}}, \bibinfo {author} {\bibfnamefont {J.~M.}\ \bibnamefont {Leedom}}, \bibinfo {author} {\bibfnamefont {M.}~\bibnamefont {Putti}},\ and\ \bibinfo {author} {\bibfnamefont {A.}~\bibnamefont {Westphal}},\ }\bibfield  {title} {\bibinfo {title} {{Gravitational Axiverse Spectroscopy: Seeing the Forest for the Axions}},\ }\href@noop {} {\  (\bibinfo {year} {2023})},\ \Eprint {https://arxiv.org/abs/2312.13431} {arXiv:2312.13431 [hep-th]} \BibitemShut {NoStop}%
\bibitem [{\citenamefont {Corb\`a}\ and\ \citenamefont {Sorbo}(2024)}]{Corba:2024tfz}%
  \BibitemOpen
  \bibfield  {author} {\bibinfo {author} {\bibfnamefont {S.~P.}\ \bibnamefont {Corb\`a}}\ and\ \bibinfo {author} {\bibfnamefont {L.}~\bibnamefont {Sorbo}},\ }\bibfield  {title} {\bibinfo {title} {{Correlated scalar perturbations and gravitational waves from axion inflation}},\ }\href@noop {} {\  (\bibinfo {year} {2024})},\ \Eprint {https://arxiv.org/abs/2403.03338} {arXiv:2403.03338 [astro-ph.CO]} \BibitemShut {NoStop}%
\bibitem [{\citenamefont {Maiti}\ \emph {et~al.}(2024)\citenamefont {Maiti}, \citenamefont {Maity},\ and\ \citenamefont {Sriramkumar}}]{Maiti:2024nhv}%
  \BibitemOpen
  \bibfield  {author} {\bibinfo {author} {\bibfnamefont {S.}~\bibnamefont {Maiti}}, \bibinfo {author} {\bibfnamefont {D.}~\bibnamefont {Maity}},\ and\ \bibinfo {author} {\bibfnamefont {L.}~\bibnamefont {Sriramkumar}},\ }\bibfield  {title} {\bibinfo {title} {{Constraining inflationary magnetogenesis and reheating via GWs in light of PTA data}},\ }\href@noop {} {\  (\bibinfo {year} {2024})},\ \Eprint {https://arxiv.org/abs/2401.01864} {arXiv:2401.01864 [gr-qc]} \BibitemShut {NoStop}%
\bibitem [{\citenamefont {Alam}\ \emph {et~al.}(2024)\citenamefont {Alam}, \citenamefont {Dutta},\ and\ \citenamefont {Jaman}}]{Alam:2024fid}%
  \BibitemOpen
  \bibfield  {author} {\bibinfo {author} {\bibfnamefont {K.}~\bibnamefont {Alam}}, \bibinfo {author} {\bibfnamefont {K.}~\bibnamefont {Dutta}},\ and\ \bibinfo {author} {\bibfnamefont {N.}~\bibnamefont {Jaman}},\ }\bibfield  {title} {\bibinfo {title} {{CMB constraints on natural inflation with gauge field production}},\ }\href {https://doi.org/10.1088/1475-7516/2024/12/015} {\bibfield  {journal} {\bibinfo  {journal} {JCAP}\ }\textbf {\bibinfo {volume} {12}},\ \bibinfo {pages} {015}},\ \Eprint {https://arxiv.org/abs/2405.10155} {arXiv:2405.10155 [astro-ph.CO]} \BibitemShut {NoStop}%
\bibitem [{\citenamefont {Domcke}\ \emph {et~al.}(2020{\natexlab{a}})\citenamefont {Domcke}, \citenamefont {Guidetti}, \citenamefont {Welling},\ and\ \citenamefont {Westphal}}]{Domcke:2020zez}%
  \BibitemOpen
  \bibfield  {author} {\bibinfo {author} {\bibfnamefont {V.}~\bibnamefont {Domcke}}, \bibinfo {author} {\bibfnamefont {V.}~\bibnamefont {Guidetti}}, \bibinfo {author} {\bibfnamefont {Y.}~\bibnamefont {Welling}},\ and\ \bibinfo {author} {\bibfnamefont {A.}~\bibnamefont {Westphal}},\ }\bibfield  {title} {\bibinfo {title} {{Resonant backreaction in axion inflation}},\ }\href {https://doi.org/10.1088/1475-7516/2020/09/009} {\bibfield  {journal} {\bibinfo  {journal} {JCAP}\ }\textbf {\bibinfo {volume} {09}},\ \bibinfo {pages} {009}},\ \Eprint {https://arxiv.org/abs/2002.02952} {arXiv:2002.02952 [astro-ph.CO]} \BibitemShut {NoStop}%
\bibitem [{\citenamefont {Peloso}\ and\ \citenamefont {Sorbo}(2023)}]{Peloso:2022ovc}%
  \BibitemOpen
  \bibfield  {author} {\bibinfo {author} {\bibfnamefont {M.}~\bibnamefont {Peloso}}\ and\ \bibinfo {author} {\bibfnamefont {L.}~\bibnamefont {Sorbo}},\ }\bibfield  {title} {\bibinfo {title} {{Instability in axion inflation with strong backreaction from gauge modes}},\ }\href {https://doi.org/10.1088/1475-7516/2023/01/038} {\bibfield  {journal} {\bibinfo  {journal} {JCAP}\ }\textbf {\bibinfo {volume} {01}},\ \bibinfo {pages} {038}},\ \Eprint {https://arxiv.org/abs/2209.08131} {arXiv:2209.08131 [astro-ph.CO]} \BibitemShut {NoStop}%
\bibitem [{\citenamefont {He}\ \emph {et~al.}(2025{\natexlab{b}})\citenamefont {He}, \citenamefont {Zhang}, \citenamefont {Fu},\ and\ \citenamefont {Guo}}]{He:2025ieo}%
  \BibitemOpen
  \bibfield  {author} {\bibinfo {author} {\bibfnamefont {J.-F.}\ \bibnamefont {He}}, \bibinfo {author} {\bibfnamefont {K.-G.}\ \bibnamefont {Zhang}}, \bibinfo {author} {\bibfnamefont {C.}~\bibnamefont {Fu}},\ and\ \bibinfo {author} {\bibfnamefont {Z.-K.}\ \bibnamefont {Guo}},\ }\bibfield  {title} {\bibinfo {title} {{Strong backreaction of gauge quanta produced during inflation}},\ }\href {https://doi.org/10.1103/PhysRevD.111.103525} {\bibfield  {journal} {\bibinfo  {journal} {Phys. Rev. D}\ }\textbf {\bibinfo {volume} {111}},\ \bibinfo {pages} {103525} (\bibinfo {year} {2025}{\natexlab{b}})},\ \Eprint {https://arxiv.org/abs/2502.13158} {arXiv:2502.13158 [hep-ph]} \BibitemShut {NoStop}%
\bibitem [{\citenamefont {Cheng}\ \emph {et~al.}(2016)\citenamefont {Cheng}, \citenamefont {Lee},\ and\ \citenamefont {Ng}}]{Cheng:2015oqa}%
  \BibitemOpen
  \bibfield  {author} {\bibinfo {author} {\bibfnamefont {S.-L.}\ \bibnamefont {Cheng}}, \bibinfo {author} {\bibfnamefont {W.}~\bibnamefont {Lee}},\ and\ \bibinfo {author} {\bibfnamefont {K.-W.}\ \bibnamefont {Ng}},\ }\bibfield  {title} {\bibinfo {title} {{Numerical study of pseudoscalar inflation with an axion-gauge field coupling}},\ }\href {https://doi.org/10.1103/PhysRevD.93.063510} {\bibfield  {journal} {\bibinfo  {journal} {Phys. Rev. D}\ }\textbf {\bibinfo {volume} {93}},\ \bibinfo {pages} {063510} (\bibinfo {year} {2016})},\ \Eprint {https://arxiv.org/abs/1508.00251} {arXiv:1508.00251 [astro-ph.CO]} \BibitemShut {NoStop}%
\bibitem [{\citenamefont {Notari}\ and\ \citenamefont {Tywoniuk}(2016)}]{Notari:2016npn}%
  \BibitemOpen
  \bibfield  {author} {\bibinfo {author} {\bibfnamefont {A.}~\bibnamefont {Notari}}\ and\ \bibinfo {author} {\bibfnamefont {K.}~\bibnamefont {Tywoniuk}},\ }\bibfield  {title} {\bibinfo {title} {{Dissipative Axial Inflation}},\ }\href {https://doi.org/10.1088/1475-7516/2016/12/038} {\bibfield  {journal} {\bibinfo  {journal} {JCAP}\ }\textbf {\bibinfo {volume} {12}},\ \bibinfo {pages} {038}},\ \Eprint {https://arxiv.org/abs/1608.06223} {arXiv:1608.06223 [hep-th]} \BibitemShut {NoStop}%
\bibitem [{\citenamefont {Dall'Agata}\ \emph {et~al.}(2020)\citenamefont {Dall'Agata}, \citenamefont {Gonz\'alez-Mart\'\i{}n}, \citenamefont {Papageorgiou},\ and\ \citenamefont {Peloso}}]{DallAgata:2019yrr}%
  \BibitemOpen
  \bibfield  {author} {\bibinfo {author} {\bibfnamefont {G.}~\bibnamefont {Dall'Agata}}, \bibinfo {author} {\bibfnamefont {S.}~\bibnamefont {Gonz\'alez-Mart\'\i{}n}}, \bibinfo {author} {\bibfnamefont {A.}~\bibnamefont {Papageorgiou}},\ and\ \bibinfo {author} {\bibfnamefont {M.}~\bibnamefont {Peloso}},\ }\bibfield  {title} {\bibinfo {title} {{Warm dark energy}},\ }\href {https://doi.org/10.1088/1475-7516/2020/08/032} {\bibfield  {journal} {\bibinfo  {journal} {JCAP}\ }\textbf {\bibinfo {volume} {08}},\ \bibinfo {pages} {032}},\ \Eprint {https://arxiv.org/abs/1912.09950} {arXiv:1912.09950 [hep-th]} \BibitemShut {NoStop}%
\bibitem [{\citenamefont {Gorbar}\ \emph {et~al.}(2021)\citenamefont {Gorbar}, \citenamefont {Schmitz}, \citenamefont {Sobol},\ and\ \citenamefont {Vilchinskii}}]{Gorbar:2021rlt}%
  \BibitemOpen
  \bibfield  {author} {\bibinfo {author} {\bibfnamefont {E.~V.}\ \bibnamefont {Gorbar}}, \bibinfo {author} {\bibfnamefont {K.}~\bibnamefont {Schmitz}}, \bibinfo {author} {\bibfnamefont {O.~O.}\ \bibnamefont {Sobol}},\ and\ \bibinfo {author} {\bibfnamefont {S.~I.}\ \bibnamefont {Vilchinskii}},\ }\bibfield  {title} {\bibinfo {title} {{Gauge-field production during axion inflation in the gradient expansion formalism}},\ }\href {https://doi.org/10.1103/PhysRevD.104.123504} {\bibfield  {journal} {\bibinfo  {journal} {Phys. Rev. D}\ }\textbf {\bibinfo {volume} {104}},\ \bibinfo {pages} {123504} (\bibinfo {year} {2021})},\ \Eprint {https://arxiv.org/abs/2109.01651} {arXiv:2109.01651 [hep-ph]} \BibitemShut {NoStop}%
\bibitem [{\citenamefont {Durrer}\ \emph {et~al.}(2023)\citenamefont {Durrer}, \citenamefont {Sobol},\ and\ \citenamefont {Vilchinskii}}]{Durrer:2023rhc}%
  \BibitemOpen
  \bibfield  {author} {\bibinfo {author} {\bibfnamefont {R.}~\bibnamefont {Durrer}}, \bibinfo {author} {\bibfnamefont {O.}~\bibnamefont {Sobol}},\ and\ \bibinfo {author} {\bibfnamefont {S.}~\bibnamefont {Vilchinskii}},\ }\bibfield  {title} {\bibinfo {title} {{Backreaction from gauge fields produced during inflation}},\ }\href {https://doi.org/10.1103/PhysRevD.108.043540} {\bibfield  {journal} {\bibinfo  {journal} {Phys. Rev. D}\ }\textbf {\bibinfo {volume} {108}},\ \bibinfo {pages} {043540} (\bibinfo {year} {2023})},\ \Eprint {https://arxiv.org/abs/2303.04583} {arXiv:2303.04583 [gr-qc]} \BibitemShut {NoStop}%
\bibitem [{\citenamefont {von Eckardstein}\ \emph {et~al.}(2023)\citenamefont {von Eckardstein}, \citenamefont {Peloso}, \citenamefont {Schmitz}, \citenamefont {Sobol},\ and\ \citenamefont {Sorbo}}]{vonEckardstein:2023gwk}%
  \BibitemOpen
  \bibfield  {author} {\bibinfo {author} {\bibfnamefont {R.}~\bibnamefont {von Eckardstein}}, \bibinfo {author} {\bibfnamefont {M.}~\bibnamefont {Peloso}}, \bibinfo {author} {\bibfnamefont {K.}~\bibnamefont {Schmitz}}, \bibinfo {author} {\bibfnamefont {O.}~\bibnamefont {Sobol}},\ and\ \bibinfo {author} {\bibfnamefont {L.}~\bibnamefont {Sorbo}},\ }\bibfield  {title} {\bibinfo {title} {{Axion inflation in the strong-backreaction regime: decay of the Anber-Sorbo solution}},\ }\href {https://doi.org/10.1007/JHEP11(2023)183} {\bibfield  {journal} {\bibinfo  {journal} {JHEP}\ }\textbf {\bibinfo {volume} {11}},\ \bibinfo {pages} {183}},\ \Eprint {https://arxiv.org/abs/2309.04254} {arXiv:2309.04254 [hep-ph]} \BibitemShut {NoStop}%
\bibitem [{\citenamefont {Iarygina}\ \emph {et~al.}(2024)\citenamefont {Iarygina}, \citenamefont {Sfakianakis}, \citenamefont {Sharma},\ and\ \citenamefont {Brandenburg}}]{Iarygina:2023mtj}%
  \BibitemOpen
  \bibfield  {author} {\bibinfo {author} {\bibfnamefont {O.}~\bibnamefont {Iarygina}}, \bibinfo {author} {\bibfnamefont {E.~I.}\ \bibnamefont {Sfakianakis}}, \bibinfo {author} {\bibfnamefont {R.}~\bibnamefont {Sharma}},\ and\ \bibinfo {author} {\bibfnamefont {A.}~\bibnamefont {Brandenburg}},\ }\bibfield  {title} {\bibinfo {title} {{Backreaction of axion-SU(2) dynamics during inflation}},\ }\href {https://doi.org/10.1088/1475-7516/2024/04/018} {\bibfield  {journal} {\bibinfo  {journal} {JCAP}\ }\textbf {\bibinfo {volume} {04}},\ \bibinfo {pages} {018}},\ \Eprint {https://arxiv.org/abs/2311.07557} {arXiv:2311.07557 [astro-ph.CO]} \BibitemShut {NoStop}%
\bibitem [{\citenamefont {Caravano}\ \emph {et~al.}(2023)\citenamefont {Caravano}, \citenamefont {Komatsu}, \citenamefont {Lozanov},\ and\ \citenamefont {Weller}}]{Caravano:2022epk}%
  \BibitemOpen
  \bibfield  {author} {\bibinfo {author} {\bibfnamefont {A.}~\bibnamefont {Caravano}}, \bibinfo {author} {\bibfnamefont {E.}~\bibnamefont {Komatsu}}, \bibinfo {author} {\bibfnamefont {K.~D.}\ \bibnamefont {Lozanov}},\ and\ \bibinfo {author} {\bibfnamefont {J.}~\bibnamefont {Weller}},\ }\bibfield  {title} {\bibinfo {title} {{Lattice simulations of axion-U(1) inflation}},\ }\href {https://doi.org/10.1103/PhysRevD.108.043504} {\bibfield  {journal} {\bibinfo  {journal} {Phys. Rev. D}\ }\textbf {\bibinfo {volume} {108}},\ \bibinfo {pages} {043504} (\bibinfo {year} {2023})},\ \Eprint {https://arxiv.org/abs/2204.12874} {arXiv:2204.12874 [astro-ph.CO]} \BibitemShut {NoStop}%
\bibitem [{\citenamefont {Galanti}\ \emph {et~al.}(2024)\citenamefont {Galanti}, \citenamefont {Conzinu}, \citenamefont {Marozzi},\ and\ \citenamefont {Santos~da Costa}}]{Galanti:2024jhw}%
  \BibitemOpen
  \bibfield  {author} {\bibinfo {author} {\bibfnamefont {D.~C.}\ \bibnamefont {Galanti}}, \bibinfo {author} {\bibfnamefont {P.}~\bibnamefont {Conzinu}}, \bibinfo {author} {\bibfnamefont {G.}~\bibnamefont {Marozzi}},\ and\ \bibinfo {author} {\bibfnamefont {S.}~\bibnamefont {Santos~da Costa}},\ }\bibfield  {title} {\bibinfo {title} {{Gauge invariant quantum backreaction in U(1) axion inflation}},\ }\href {https://doi.org/10.1103/PhysRevD.110.123510} {\bibfield  {journal} {\bibinfo  {journal} {Phys. Rev. D}\ }\textbf {\bibinfo {volume} {110}},\ \bibinfo {pages} {123510} (\bibinfo {year} {2024})},\ \Eprint {https://arxiv.org/abs/2406.19960} {arXiv:2406.19960 [gr-qc]} \BibitemShut {NoStop}%
\bibitem [{\citenamefont {Caravano}\ and\ \citenamefont {Peloso}(2024)}]{Caravano:2024xsb}%
  \BibitemOpen
  \bibfield  {author} {\bibinfo {author} {\bibfnamefont {A.}~\bibnamefont {Caravano}}\ and\ \bibinfo {author} {\bibfnamefont {M.}~\bibnamefont {Peloso}},\ }\bibfield  {title} {\bibinfo {title} {{Unveiling the nonlinear dynamics of a rolling axion during inflation}},\ }\href@noop {} {\  (\bibinfo {year} {2024})},\ \Eprint {https://arxiv.org/abs/2407.13405} {arXiv:2407.13405 [astro-ph.CO]} \BibitemShut {NoStop}%
\bibitem [{\citenamefont {Figueroa}\ \emph {et~al.}(2024{\natexlab{b}})\citenamefont {Figueroa}, \citenamefont {Lizarraga}, \citenamefont {Loayza}, \citenamefont {Urio},\ and\ \citenamefont {Urrestilla}}]{Figueroa:2024rkr}%
  \BibitemOpen
  \bibfield  {author} {\bibinfo {author} {\bibfnamefont {D.~G.}\ \bibnamefont {Figueroa}}, \bibinfo {author} {\bibfnamefont {J.}~\bibnamefont {Lizarraga}}, \bibinfo {author} {\bibfnamefont {N.}~\bibnamefont {Loayza}}, \bibinfo {author} {\bibfnamefont {A.}~\bibnamefont {Urio}},\ and\ \bibinfo {author} {\bibfnamefont {J.}~\bibnamefont {Urrestilla}},\ }\bibfield  {title} {\bibinfo {title} {{The non-linear dynamics of axion inflation: a detailed lattice study}},\ }\href@noop {} {\  (\bibinfo {year} {2024}{\natexlab{b}})},\ \Eprint {https://arxiv.org/abs/2411.16368} {arXiv:2411.16368 [astro-ph.CO]} \BibitemShut {NoStop}%
\bibitem [{\citenamefont {Sharma}\ \emph {et~al.}(2024)\citenamefont {Sharma}, \citenamefont {Brandenburg}, \citenamefont {Subramanian},\ and\ \citenamefont {Vikman}}]{Sharma:2024nfu}%
  \BibitemOpen
  \bibfield  {author} {\bibinfo {author} {\bibfnamefont {R.}~\bibnamefont {Sharma}}, \bibinfo {author} {\bibfnamefont {A.}~\bibnamefont {Brandenburg}}, \bibinfo {author} {\bibfnamefont {K.}~\bibnamefont {Subramanian}},\ and\ \bibinfo {author} {\bibfnamefont {A.}~\bibnamefont {Vikman}},\ }\bibfield  {title} {\bibinfo {title} {{Lattice simulations of axion-U(1) inflation: gravitational waves, magnetic fields, and black holes}},\ }\href@noop {} {\  (\bibinfo {year} {2024})},\ \Eprint {https://arxiv.org/abs/2411.04854} {arXiv:2411.04854 [astro-ph.CO]} \BibitemShut {NoStop}%
\bibitem [{\citenamefont {Lizarraga}\ \emph {et~al.}(2025)\citenamefont {Lizarraga}, \citenamefont {L\'opez-Mediavilla},\ and\ \citenamefont {Urio}}]{Lizarraga:2025aiw}%
  \BibitemOpen
  \bibfield  {author} {\bibinfo {author} {\bibfnamefont {J.}~\bibnamefont {Lizarraga}}, \bibinfo {author} {\bibfnamefont {C.}~\bibnamefont {L\'opez-Mediavilla}},\ and\ \bibinfo {author} {\bibfnamefont {A.}~\bibnamefont {Urio}},\ }\bibfield  {title} {\bibinfo {title} {{Comparative study of the strong backreaction regime in axion inflation: the effect of the potential}},\ }\href@noop {} {\  (\bibinfo {year} {2025})},\ \Eprint {https://arxiv.org/abs/2505.19950} {arXiv:2505.19950 [astro-ph.CO]} \BibitemShut {NoStop}%
\bibitem [{\citenamefont {Bhattacharya}\ \emph {et~al.}(2025)\citenamefont {Bhattacharya}, \citenamefont {Fasiello}, \citenamefont {Papageorgiou},\ and\ \citenamefont {Dimastrogiovanni}}]{Bhattacharya:2025guc}%
  \BibitemOpen
  \bibfield  {author} {\bibinfo {author} {\bibfnamefont {S.}~\bibnamefont {Bhattacharya}}, \bibinfo {author} {\bibfnamefont {M.}~\bibnamefont {Fasiello}}, \bibinfo {author} {\bibfnamefont {A.}~\bibnamefont {Papageorgiou}},\ and\ \bibinfo {author} {\bibfnamefont {E.}~\bibnamefont {Dimastrogiovanni}},\ }\bibfield  {title} {\bibinfo {title} {{On the prospects of thermalization of axion-SU(2) inflation}},\ }\href@noop {} {\  (\bibinfo {year} {2025})},\ \Eprint {https://arxiv.org/abs/2506.11853} {arXiv:2506.11853 [astro-ph.CO]} \BibitemShut {NoStop}%
\bibitem [{\citenamefont {Banks}\ \emph {et~al.}(2003)\citenamefont {Banks}, \citenamefont {Dine}, \citenamefont {Fox},\ and\ \citenamefont {Gorbatov}}]{Banks:2003sx}%
  \BibitemOpen
  \bibfield  {author} {\bibinfo {author} {\bibfnamefont {T.}~\bibnamefont {Banks}}, \bibinfo {author} {\bibfnamefont {M.}~\bibnamefont {Dine}}, \bibinfo {author} {\bibfnamefont {P.~J.}\ \bibnamefont {Fox}},\ and\ \bibinfo {author} {\bibfnamefont {E.}~\bibnamefont {Gorbatov}},\ }\bibfield  {title} {\bibinfo {title} {{On the possibility of large axion decay constants}},\ }\href {https://doi.org/10.1088/1475-7516/2003/06/001} {\bibfield  {journal} {\bibinfo  {journal} {JCAP}\ }\textbf {\bibinfo {volume} {06}},\ \bibinfo {pages} {001}},\ \Eprint {https://arxiv.org/abs/hep-th/0303252} {arXiv:hep-th/0303252} \BibitemShut {NoStop}%
\bibitem [{\citenamefont {Arkani-Hamed}\ \emph {et~al.}(2007)\citenamefont {Arkani-Hamed}, \citenamefont {Motl}, \citenamefont {Nicolis},\ and\ \citenamefont {Vafa}}]{Arkani-Hamed:2006emk}%
  \BibitemOpen
  \bibfield  {author} {\bibinfo {author} {\bibfnamefont {N.}~\bibnamefont {Arkani-Hamed}}, \bibinfo {author} {\bibfnamefont {L.}~\bibnamefont {Motl}}, \bibinfo {author} {\bibfnamefont {A.}~\bibnamefont {Nicolis}},\ and\ \bibinfo {author} {\bibfnamefont {C.}~\bibnamefont {Vafa}},\ }\bibfield  {title} {\bibinfo {title} {{The String landscape, black holes and gravity as the weakest force}},\ }\href {https://doi.org/10.1088/1126-6708/2007/06/060} {\bibfield  {journal} {\bibinfo  {journal} {JHEP}\ }\textbf {\bibinfo {volume} {06}},\ \bibinfo {pages} {060}},\ \Eprint {https://arxiv.org/abs/hep-th/0601001} {arXiv:hep-th/0601001} \BibitemShut {NoStop}%
\bibitem [{\citenamefont {Svrcek}\ and\ \citenamefont {Witten}(2006)}]{Svrcek:2006yi}%
  \BibitemOpen
  \bibfield  {author} {\bibinfo {author} {\bibfnamefont {P.}~\bibnamefont {Svrcek}}\ and\ \bibinfo {author} {\bibfnamefont {E.}~\bibnamefont {Witten}},\ }\bibfield  {title} {\bibinfo {title} {{Axions In String Theory}},\ }\href {https://doi.org/10.1088/1126-6708/2006/06/051} {\bibfield  {journal} {\bibinfo  {journal} {JHEP}\ }\textbf {\bibinfo {volume} {06}},\ \bibinfo {pages} {051}},\ \Eprint {https://arxiv.org/abs/hep-th/0605206} {arXiv:hep-th/0605206} \BibitemShut {NoStop}%
\bibitem [{\citenamefont {Silverstein}\ and\ \citenamefont {Westphal}(2008)}]{Silverstein:2008sg}%
  \BibitemOpen
  \bibfield  {author} {\bibinfo {author} {\bibfnamefont {E.}~\bibnamefont {Silverstein}}\ and\ \bibinfo {author} {\bibfnamefont {A.}~\bibnamefont {Westphal}},\ }\bibfield  {title} {\bibinfo {title} {{Monodromy in the CMB: Gravity Waves and String Inflation}},\ }\href {https://doi.org/10.1103/PhysRevD.78.106003} {\bibfield  {journal} {\bibinfo  {journal} {Phys. Rev. D}\ }\textbf {\bibinfo {volume} {78}},\ \bibinfo {pages} {106003} (\bibinfo {year} {2008})},\ \Eprint {https://arxiv.org/abs/0803.3085} {arXiv:0803.3085 [hep-th]} \BibitemShut {NoStop}%
\bibitem [{\citenamefont {McAllister}\ \emph {et~al.}(2010)\citenamefont {McAllister}, \citenamefont {Silverstein},\ and\ \citenamefont {Westphal}}]{McAllister:2008hb}%
  \BibitemOpen
  \bibfield  {author} {\bibinfo {author} {\bibfnamefont {L.}~\bibnamefont {McAllister}}, \bibinfo {author} {\bibfnamefont {E.}~\bibnamefont {Silverstein}},\ and\ \bibinfo {author} {\bibfnamefont {A.}~\bibnamefont {Westphal}},\ }\bibfield  {title} {\bibinfo {title} {{Gravity Waves and Linear Inflation from Axion Monodromy}},\ }\href {https://doi.org/10.1103/PhysRevD.82.046003} {\bibfield  {journal} {\bibinfo  {journal} {Phys. Rev. D}\ }\textbf {\bibinfo {volume} {82}},\ \bibinfo {pages} {046003} (\bibinfo {year} {2010})},\ \Eprint {https://arxiv.org/abs/0808.0706} {arXiv:0808.0706 [hep-th]} \BibitemShut {NoStop}%
\bibitem [{\citenamefont {Kobayashi}\ \emph {et~al.}(2016)\citenamefont {Kobayashi}, \citenamefont {Oikawa},\ and\ \citenamefont {Otsuka}}]{Kobayashi:2015aaa}%
  \BibitemOpen
  \bibfield  {author} {\bibinfo {author} {\bibfnamefont {T.}~\bibnamefont {Kobayashi}}, \bibinfo {author} {\bibfnamefont {A.}~\bibnamefont {Oikawa}},\ and\ \bibinfo {author} {\bibfnamefont {H.}~\bibnamefont {Otsuka}},\ }\bibfield  {title} {\bibinfo {title} {{New potentials for string axion inflation}},\ }\href {https://doi.org/10.1103/PhysRevD.93.083508} {\bibfield  {journal} {\bibinfo  {journal} {Phys. Rev. D}\ }\textbf {\bibinfo {volume} {93}},\ \bibinfo {pages} {083508} (\bibinfo {year} {2016})},\ \Eprint {https://arxiv.org/abs/1510.08768} {arXiv:1510.08768 [hep-ph]} \BibitemShut {NoStop}%
\bibitem [{\citenamefont {Cabo~Bizet}\ \emph {et~al.}(2016)\citenamefont {Cabo~Bizet}, \citenamefont {Loaiza-Brito},\ and\ \citenamefont {Zavala}}]{CaboBizet:2016uzv}%
  \BibitemOpen
  \bibfield  {author} {\bibinfo {author} {\bibfnamefont {N.}~\bibnamefont {Cabo~Bizet}}, \bibinfo {author} {\bibfnamefont {O.}~\bibnamefont {Loaiza-Brito}},\ and\ \bibinfo {author} {\bibfnamefont {I.}~\bibnamefont {Zavala}},\ }\bibfield  {title} {\bibinfo {title} {{Mirror quintic vacua: hierarchies and inflation}},\ }\href {https://doi.org/10.1007/JHEP10(2016)082} {\bibfield  {journal} {\bibinfo  {journal} {JHEP}\ }\textbf {\bibinfo {volume} {10}},\ \bibinfo {pages} {082}},\ \Eprint {https://arxiv.org/abs/1605.03974} {arXiv:1605.03974 [hep-th]} \BibitemShut {NoStop}%
\bibitem [{\citenamefont {Carrasco}\ \emph {et~al.}(2015)\citenamefont {Carrasco}, \citenamefont {Kallosh},\ and\ \citenamefont {Linde}}]{Carrasco:2015pla}%
  \BibitemOpen
  \bibfield  {author} {\bibinfo {author} {\bibfnamefont {J.~J.~M.}\ \bibnamefont {Carrasco}}, \bibinfo {author} {\bibfnamefont {R.}~\bibnamefont {Kallosh}},\ and\ \bibinfo {author} {\bibfnamefont {A.}~\bibnamefont {Linde}},\ }\bibfield  {title} {\bibinfo {title} {{$\alpha $-Attractors: Planck, LHC and Dark Energy}},\ }\href {https://doi.org/10.1007/JHEP10(2015)147} {\bibfield  {journal} {\bibinfo  {journal} {JHEP}\ }\textbf {\bibinfo {volume} {10}},\ \bibinfo {pages} {147}},\ \Eprint {https://arxiv.org/abs/1506.01708} {arXiv:1506.01708 [hep-th]} \BibitemShut {NoStop}%
\bibitem [{\citenamefont {Kallosh}\ and\ \citenamefont {Linde}(2022)}]{Kallosh:2022feu}%
  \BibitemOpen
  \bibfield  {author} {\bibinfo {author} {\bibfnamefont {R.}~\bibnamefont {Kallosh}}\ and\ \bibinfo {author} {\bibfnamefont {A.}~\bibnamefont {Linde}},\ }\bibfield  {title} {\bibinfo {title} {{Polynomial \ensuremath{\alpha}-attractors}},\ }\href {https://doi.org/10.1088/1475-7516/2022/04/017} {\bibfield  {journal} {\bibinfo  {journal} {JCAP}\ }\textbf {\bibinfo {volume} {04}}\bibfield  {number} {\bibinfo  {number} { (04)},\ \bibinfo {pages} {017}},\ }\Eprint {https://arxiv.org/abs/2202.06492} {arXiv:2202.06492 [astro-ph.CO]} \BibitemShut {NoStop}%
\bibitem [{\citenamefont {\"Ozsoy}(2021)}]{Ozsoy:2020ccy}%
  \BibitemOpen
  \bibfield  {author} {\bibinfo {author} {\bibfnamefont {O.}~\bibnamefont {\"Ozsoy}},\ }\bibfield  {title} {\bibinfo {title} {{Synthetic Gravitational Waves from a Rolling Axion Monodromy}},\ }\href {https://doi.org/10.1088/1475-7516/2021/04/040} {\bibfield  {journal} {\bibinfo  {journal} {JCAP}\ }\textbf {\bibinfo {volume} {04}},\ \bibinfo {pages} {040}},\ \Eprint {https://arxiv.org/abs/2005.10280} {arXiv:2005.10280 [astro-ph.CO]} \BibitemShut {NoStop}%
\bibitem [{\citenamefont {Seto}\ and\ \citenamefont {Taruya}(2007)}]{Seto:2007tn}%
  \BibitemOpen
  \bibfield  {author} {\bibinfo {author} {\bibfnamefont {N.}~\bibnamefont {Seto}}\ and\ \bibinfo {author} {\bibfnamefont {A.}~\bibnamefont {Taruya}},\ }\bibfield  {title} {\bibinfo {title} {{Measuring a Parity Violation Signature in the Early Universe via Ground-based Laser Interferometers}},\ }\href {https://doi.org/10.1103/PhysRevLett.99.121101} {\bibfield  {journal} {\bibinfo  {journal} {Phys. Rev. Lett.}\ }\textbf {\bibinfo {volume} {99}},\ \bibinfo {pages} {121101} (\bibinfo {year} {2007})},\ \Eprint {https://arxiv.org/abs/0707.0535} {arXiv:0707.0535 [astro-ph]} \BibitemShut {NoStop}%
\bibitem [{\citenamefont {Gluscevic}\ and\ \citenamefont {Kamionkowski}(2010)}]{Gluscevic:2010vv}%
  \BibitemOpen
  \bibfield  {author} {\bibinfo {author} {\bibfnamefont {V.}~\bibnamefont {Gluscevic}}\ and\ \bibinfo {author} {\bibfnamefont {M.}~\bibnamefont {Kamionkowski}},\ }\bibfield  {title} {\bibinfo {title} {{Testing Parity-Violating Mechanisms with Cosmic Microwave Background Experiments}},\ }\href {https://doi.org/10.1103/PhysRevD.81.123529} {\bibfield  {journal} {\bibinfo  {journal} {Phys. Rev. D}\ }\textbf {\bibinfo {volume} {81}},\ \bibinfo {pages} {123529} (\bibinfo {year} {2010})},\ \Eprint {https://arxiv.org/abs/1002.1308} {arXiv:1002.1308 [astro-ph.CO]} \BibitemShut {NoStop}%
\bibitem [{\citenamefont {Smith}\ and\ \citenamefont {Caldwell}(2017)}]{Smith:2016jqs}%
  \BibitemOpen
  \bibfield  {author} {\bibinfo {author} {\bibfnamefont {T.~L.}\ \bibnamefont {Smith}}\ and\ \bibinfo {author} {\bibfnamefont {R.}~\bibnamefont {Caldwell}},\ }\bibfield  {title} {\bibinfo {title} {{Sensitivity to a Frequency-Dependent Circular Polarization in an Isotropic Stochastic Gravitational Wave Background}},\ }\href {https://doi.org/10.1103/PhysRevD.95.044036} {\bibfield  {journal} {\bibinfo  {journal} {Phys. Rev. D}\ }\textbf {\bibinfo {volume} {95}},\ \bibinfo {pages} {044036} (\bibinfo {year} {2017})},\ \Eprint {https://arxiv.org/abs/1609.05901} {arXiv:1609.05901 [gr-qc]} \BibitemShut {NoStop}%
\bibitem [{\citenamefont {Domcke}\ \emph {et~al.}(2020{\natexlab{b}})\citenamefont {Domcke}, \citenamefont {Garcia-Bellido}, \citenamefont {Peloso}, \citenamefont {Pieroni}, \citenamefont {Ricciardone}, \citenamefont {Sorbo},\ and\ \citenamefont {Tasinato}}]{Domcke:2019zls}%
  \BibitemOpen
  \bibfield  {author} {\bibinfo {author} {\bibfnamefont {V.}~\bibnamefont {Domcke}}, \bibinfo {author} {\bibfnamefont {J.}~\bibnamefont {Garcia-Bellido}}, \bibinfo {author} {\bibfnamefont {M.}~\bibnamefont {Peloso}}, \bibinfo {author} {\bibfnamefont {M.}~\bibnamefont {Pieroni}}, \bibinfo {author} {\bibfnamefont {A.}~\bibnamefont {Ricciardone}}, \bibinfo {author} {\bibfnamefont {L.}~\bibnamefont {Sorbo}},\ and\ \bibinfo {author} {\bibfnamefont {G.}~\bibnamefont {Tasinato}},\ }\bibfield  {title} {\bibinfo {title} {{Measuring the net circular polarization of the stochastic gravitational wave background with interferometers}},\ }\href {https://doi.org/10.1088/1475-7516/2020/05/028} {\bibfield  {journal} {\bibinfo  {journal} {JCAP}\ }\textbf {\bibinfo {volume} {05}},\ \bibinfo {pages} {028}},\ \Eprint {https://arxiv.org/abs/1910.08052} {arXiv:1910.08052 [astro-ph.CO]} \BibitemShut {NoStop}%
\bibitem [{\citenamefont {Caprini}\ and\ \citenamefont {Figueroa}(2018)}]{Caprini:2018mtu}%
  \BibitemOpen
  \bibfield  {author} {\bibinfo {author} {\bibfnamefont {C.}~\bibnamefont {Caprini}}\ and\ \bibinfo {author} {\bibfnamefont {D.~G.}\ \bibnamefont {Figueroa}},\ }\bibfield  {title} {\bibinfo {title} {{Cosmological Backgrounds of Gravitational Waves}},\ }\href {https://doi.org/10.1088/1361-6382/aac608} {\bibfield  {journal} {\bibinfo  {journal} {Class. Quant. Grav.}\ }\textbf {\bibinfo {volume} {35}},\ \bibinfo {pages} {163001} (\bibinfo {year} {2018})},\ \Eprint {https://arxiv.org/abs/1801.04268} {arXiv:1801.04268 [astro-ph.CO]} \BibitemShut {NoStop}%
\bibitem [{\citenamefont {Anber}\ and\ \citenamefont {Sorbo}(2010)}]{Anber:2009ua}%
  \BibitemOpen
  \bibfield  {author} {\bibinfo {author} {\bibfnamefont {M.~M.}\ \bibnamefont {Anber}}\ and\ \bibinfo {author} {\bibfnamefont {L.}~\bibnamefont {Sorbo}},\ }\bibfield  {title} {\bibinfo {title} {{Naturally inflating on steep potentials through electromagnetic dissipation}},\ }\href {https://doi.org/10.1103/PhysRevD.81.043534} {\bibfield  {journal} {\bibinfo  {journal} {Phys. Rev. D}\ }\textbf {\bibinfo {volume} {81}},\ \bibinfo {pages} {043534} (\bibinfo {year} {2010})},\ \Eprint {https://arxiv.org/abs/0908.4089} {arXiv:0908.4089 [hep-th]} \BibitemShut {NoStop}%
\bibitem [{\citenamefont {Barnaby}\ and\ \citenamefont {Shandera}(2012)}]{Barnaby:2011pe}%
  \BibitemOpen
  \bibfield  {author} {\bibinfo {author} {\bibfnamefont {N.}~\bibnamefont {Barnaby}}\ and\ \bibinfo {author} {\bibfnamefont {S.}~\bibnamefont {Shandera}},\ }\bibfield  {title} {\bibinfo {title} {{Feeding your Inflaton: Non-Gaussian Signatures of Interaction Structure}},\ }\href {https://doi.org/10.1088/1475-7516/2012/01/034} {\bibfield  {journal} {\bibinfo  {journal} {JCAP}\ }\textbf {\bibinfo {volume} {01}},\ \bibinfo {pages} {034}},\ \Eprint {https://arxiv.org/abs/1109.2985} {arXiv:1109.2985 [astro-ph.CO]} \BibitemShut {NoStop}%
\bibitem [{\citenamefont {Jamieson}\ \emph {et~al.}(2025)\citenamefont {Jamieson}, \citenamefont {Caravano},\ and\ \citenamefont {Komatsu}}]{Jamieson:2025ngu}%
  \BibitemOpen
  \bibfield  {author} {\bibinfo {author} {\bibfnamefont {D.}~\bibnamefont {Jamieson}}, \bibinfo {author} {\bibfnamefont {A.}~\bibnamefont {Caravano}},\ and\ \bibinfo {author} {\bibfnamefont {E.}~\bibnamefont {Komatsu}},\ }\bibfield  {title} {\bibinfo {title} {{Primordial Power Spectrum and Bispectrum from Lattice Simulations of Axion-U(1) Inflation}},\ }\href@noop {} {\  (\bibinfo {year} {2025})},\ \Eprint {https://arxiv.org/abs/2507.22285} {arXiv:2507.22285 [astro-ph.CO]} \BibitemShut {NoStop}%
\bibitem [{\citenamefont {Caravano}\ \emph {et~al.}(2022)\citenamefont {Caravano}, \citenamefont {Komatsu}, \citenamefont {Lozanov},\ and\ \citenamefont {Weller}}]{Caravano:2021bfn}%
  \BibitemOpen
  \bibfield  {author} {\bibinfo {author} {\bibfnamefont {A.}~\bibnamefont {Caravano}}, \bibinfo {author} {\bibfnamefont {E.}~\bibnamefont {Komatsu}}, \bibinfo {author} {\bibfnamefont {K.~D.}\ \bibnamefont {Lozanov}},\ and\ \bibinfo {author} {\bibfnamefont {J.}~\bibnamefont {Weller}},\ }\bibfield  {title} {\bibinfo {title} {{Lattice simulations of Abelian gauge fields coupled to axions during inflation}},\ }\href {https://doi.org/10.1103/PhysRevD.105.123530} {\bibfield  {journal} {\bibinfo  {journal} {Phys. Rev. D}\ }\textbf {\bibinfo {volume} {105}},\ \bibinfo {pages} {123530} (\bibinfo {year} {2022})},\ \Eprint {https://arxiv.org/abs/2110.10695} {arXiv:2110.10695 [astro-ph.CO]} \BibitemShut {NoStop}%
\bibitem [{\citenamefont {Figueroa}\ \emph {et~al.}(2023)\citenamefont {Figueroa}, \citenamefont {Lizarraga}, \citenamefont {Urio},\ and\ \citenamefont {Urrestilla}}]{Figueroa:2023oxc}%
  \BibitemOpen
  \bibfield  {author} {\bibinfo {author} {\bibfnamefont {D.~G.}\ \bibnamefont {Figueroa}}, \bibinfo {author} {\bibfnamefont {J.}~\bibnamefont {Lizarraga}}, \bibinfo {author} {\bibfnamefont {A.}~\bibnamefont {Urio}},\ and\ \bibinfo {author} {\bibfnamefont {J.}~\bibnamefont {Urrestilla}},\ }\bibfield  {title} {\bibinfo {title} {{Strong Backreaction Regime in Axion Inflation}},\ }\href {https://doi.org/10.1103/PhysRevLett.131.151003} {\bibfield  {journal} {\bibinfo  {journal} {Phys. Rev. Lett.}\ }\textbf {\bibinfo {volume} {131}},\ \bibinfo {pages} {151003} (\bibinfo {year} {2023})},\ \Eprint {https://arxiv.org/abs/2303.17436} {arXiv:2303.17436 [astro-ph.CO]} \BibitemShut {NoStop}%
\end{thebibliography}%

\appendix

\end{document}